# YouTube Chatter: Understanding Online Comments Discourse on Misinformative and Political YouTube Videos

Aarash Heydari[1], Janny Zhang[1], Shaan Appel[2], and Xinyi Wu[3]
Professor Gireeja Ranade

## Abstract

We conduct a preliminary analysis of comments on political YouTube content containing misinformation in comparison to comments on trustworthy or apolitical videos, labelling the bias and factual ratings of our channels according to Media Bias Fact Check where applicable. One of our most interesting discoveries is that especially-polarized or misinformative political channels (Left-Bias, Right-Bias, PragerU, Conspiracy-Pseudoscience, and Questionable Source) generate 7.5x more comments per view and 10.42x more replies per view than apolitical or Pro-Science channels; in particular, Conspiracy-Pseudoscience and Questionable Sources generate 8.3x more comments per view and 11.0x more replies per view than apolitical and Pro-Science channels. We also compared average thread lengths, average comment lengths, and profanity rates across channels, and present simple machine learning classifiers for predicting the bias category of a video based on these statistics.

## Motivation

Many social media platforms have recently seen an increase in the spread of misinformation, defined as information which is misleading or factually incorrect. This spread was well documented in the wake of the 2016 US presidential election, and has colloquially been labelled as the spread of "fake news."

Inflammatory articles and stories often incorporate video content, and their comment sections provide an approximate measure of how people react to a given story. Despite this, comment threads on YouTube videos have not been studied in the context of fake news as much as on other social media platforms such as Facebook and Reddit, likely since YouTube emphasizes hosting videos whereas websites like Reddit focus on encouraging discourse. Our aim is to understand the difference in YouTube comment responses to misinformative and factual videos, as well as to political and apolitical videos.

*Note*: We avoid using the term "disinformation" in our work because it implies that said information is purposefully spreading misleading or factually incorrect information for political, profitable, or other reasons, and we cannot precisely determine the intent of our sources. Like many in the field, we also avoid using the term "fake news" to refrain from bringing its politically charged connotations into our work.

## Data

Because determining bias and identifying misinformation is not a fully objective process, we rely heavily on labels provided by Media Bias Fact Check (https://mediabiasfactcheck.com/about/) to categorize the sources for our collected data. Media Bias Fact Check (MBFC) is a source of independent critique which uses an explicitly defined methodology to label both the bias and the factual trustworthiness of many media sources.

### Channel Selection

We selected 15 YouTube channels for analysis in this work. They are primarily news sources for political content and come from a variety of perspectives. We defer to MBFC to categorize bias and classify the trustworthiness (Factual Rating) of our sources. The descriptions for each of the categories and factual ratings of our data are in item 1 of our appendix.

The biases and factual ratings of our sources are depicted in Figure 1.

**Figure 1: A graphic depicting the MBFC categorizations and MBFC factual ratings (where applicable) for the fourteen channels we scraped. We made some categories of our own, marked as [unofficial], for later reference.**

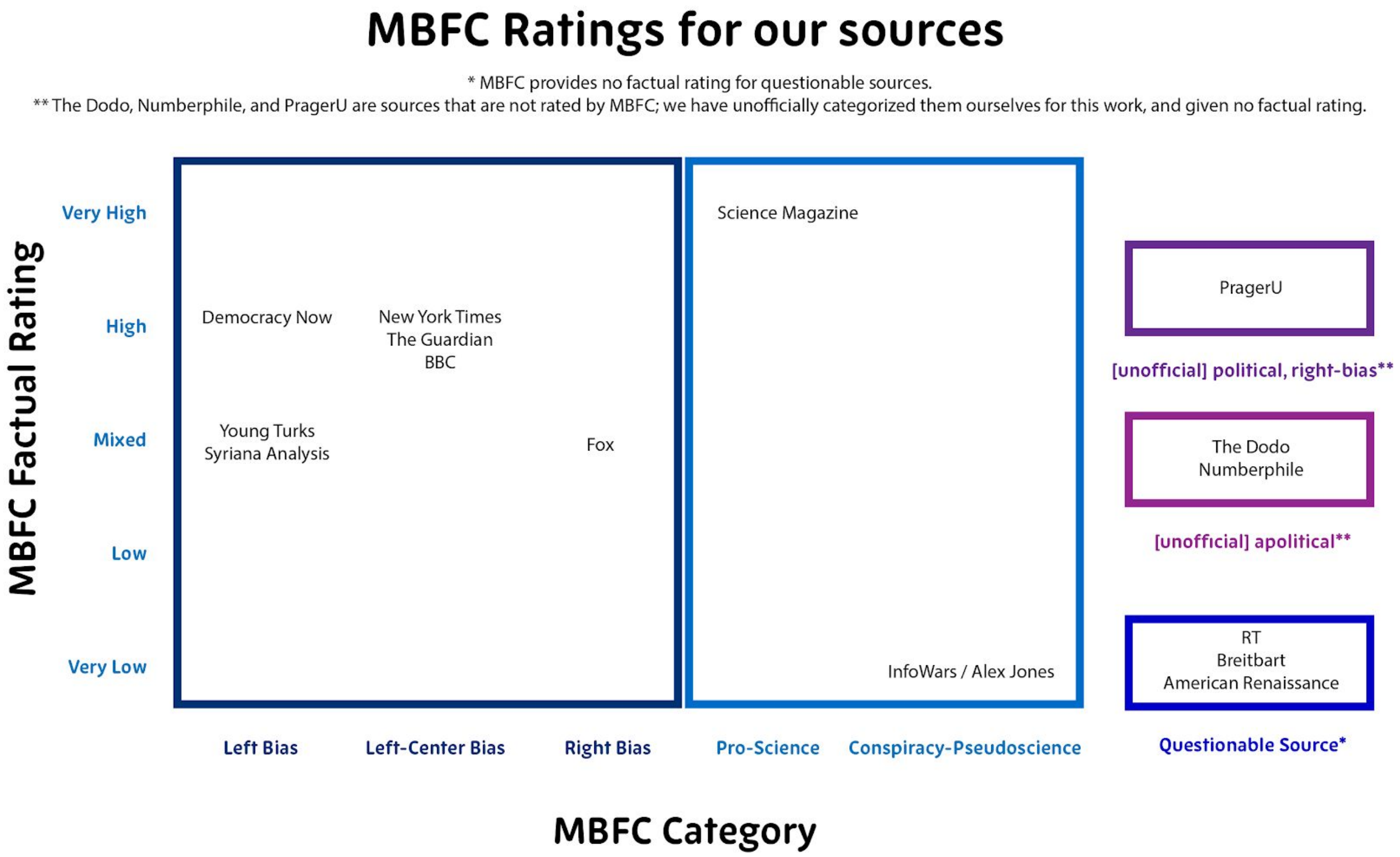

We chose our channels to cover a wide range of media. We included popular news sources such as Fox, BBC, and the New York Times; “Questionable Sources” such as Breitbart, American Renaissance, and Infowars; and sources less-focused on politics such as The Dodo, Numberphile and Science Magazine to act as a control group. Note that even channels with a high MBFC Factual Rating can be considered biased.

Six of our sources support left-leaning views (New York Times, The Guardian, BBC, Democracy Now, Young Turks, Syriana Analysis); six of our sources support right-leaning views (Fox, InfoWars/ Alex Jones, RT, Breitbart, American Renaissance, PragerU); and 3 of our sources are relatively apolitical (The Dodo, Numberphile, and Science Magazine). All of these categorizations save for The Dodo, Numberphile, and PragerU are by MBFC; we unofficially labelled the last three channels based on Wikipedia descriptions and separate those from MBFC-labelled sources in our analyses. For details on each channel, see item 2 in the appendix.

Figure 2 displays the average number of views per video for each of the different bias groups to provide context for our later work.

**Figure 2: Views across all sampled videos for the YouTube channels, normalized by the number of videos per group. Grouped visually by category and factual rating. Views measured in thousands. A chart of the views by channel can be found in the appendix, item 3.**

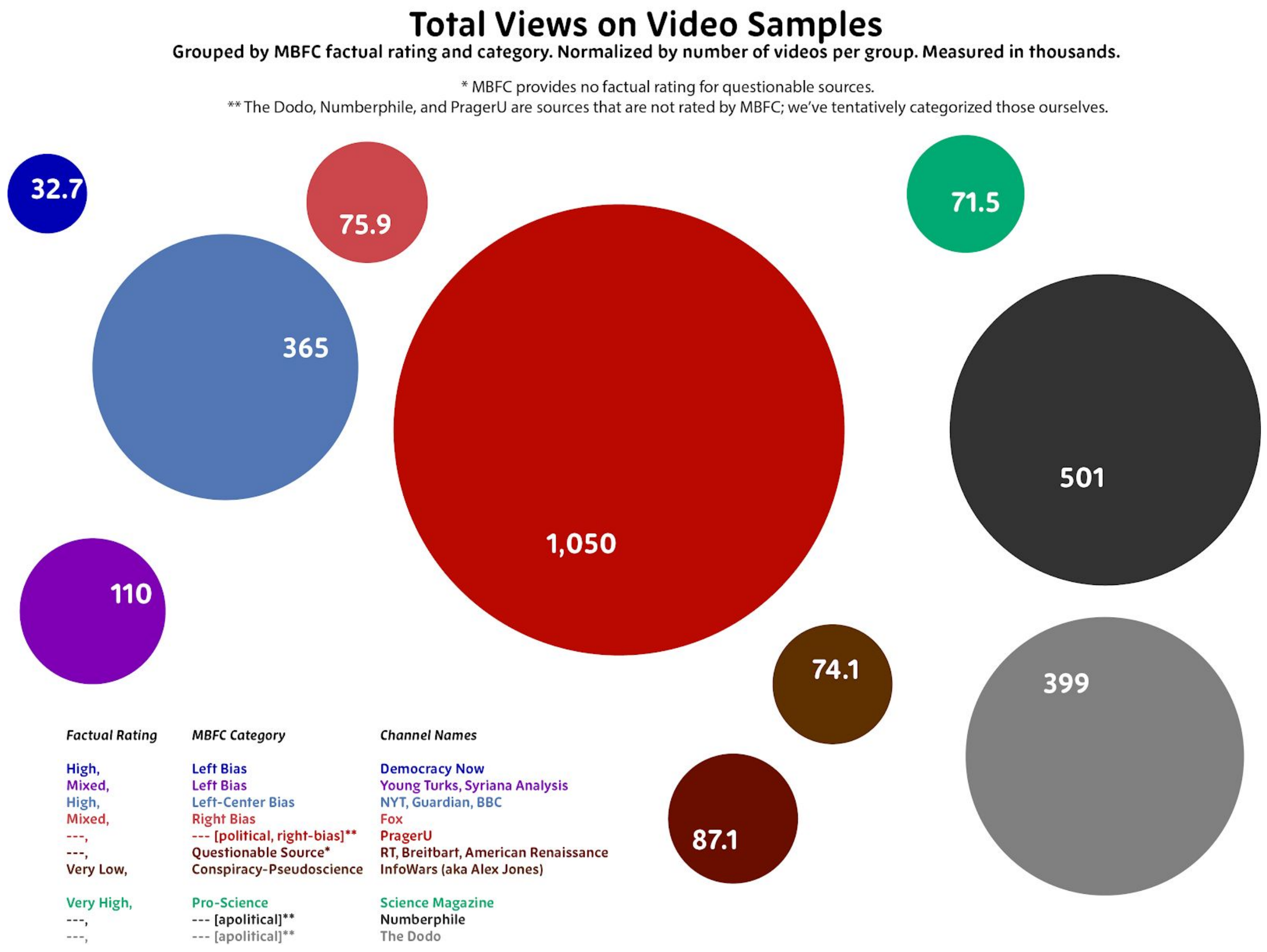

*Note*: Among our chosen channels, we see significant differences in viewership, which could add a confounding variable to our later analyses. Channels with higher view-counts often have lower engagement rates, so if one of our categories has many highly-viewed channels, that category may be erroneously labelled as having less engagement than another category.

As this is an exploratory work, we have included a Released Datasets & Code section to allow others to continue and build off our work.

**Data Collection**

We used the YouTube API found at https://developers.google.com/youtube/v3/docs/ to collect comments. For each of the channels we selected, we tried to collect the 200 most recent videos which were more than two weeks old. We ignored videos less than two weeks old in order to give them enough time to saturate with comments. The only exception to the two-week rule is the InfoWars/Alex Jones data subset (150 videos), which was deleted from YouTube before we began enforcing it. At the time we collected the American Renaissance dataset, they only had 172 videos which fit our time-step requirements. All our further analyses are done on the (up to) 200 most recent videos per channel.

For each video, we retrieve all video metadata, all comments on the video (**direct comments**), and all replies to these comments (**replies**). The **video metadata** consists of view count, like count, dislike count, favorite count, publish date, date accessed by the scraper, the publisher's channel title, and the duration of the video. Each **comment** (which includes both direct comments and replies) also has its own metadata consisting of the author's display name, timestamp of the comment, and a like count. For more details on comment structure and how the metadata is gathered, please refer to the YouTube API we linked earlier in this section.

## Released Datasets & Code

**Datasets**

To access 14 of our 15 analyzed datasets, please request access to this Google Drive folder. We have released all but our Infowars/ Alex Jones dataset.

Each dataset, save for American Renaissance, has information on 201 videos. The American Renaissance dataset has information on 172 videos.

**Code**

Find the code we used for comments-scraping and analysis at this GitHub link. You can use it to generate your own YouTube comment data from whichever channels you like. Our analyses follow in the Results section. Please cite this work if you use the code.

## Results

**Comment Engagement For Political Content:**

*Summary*

Our data suggests that political content generates significantly more engagement via comments than apolitical or pro-science content. It also suggests that biased content gets more comment engagement than moderate content, through a quick comparison between Left-Bias and Left-Center channels. On channels labelled as Conspiracy-Pseudoscience and Questionable Sources, we see an even greater engagement per user. Political or biased content also tends to have more profane comments than apolitical or less-biased content.

*Comments per view (CPV)*

Some channels were significantly more popular than others, with disparities up to a factor of 200 between the total views on the channels we chose (Figure 2). We wanted to see how well our different channels solicited responses from their viewers, and measure this by seeing how many **comments per view (CPV)** each channel received. We measure CPV by channel by going through each video in a channel and finding how many comments that video received per view, then averaging that ratio over all the videos in the channel (Figure 3A). To measure CPV by category, we take the CPV's of each of the channels in that category and average those (Figure 3B).

**Figure 3A: Total CPV, organized by channel, within our dataset.**

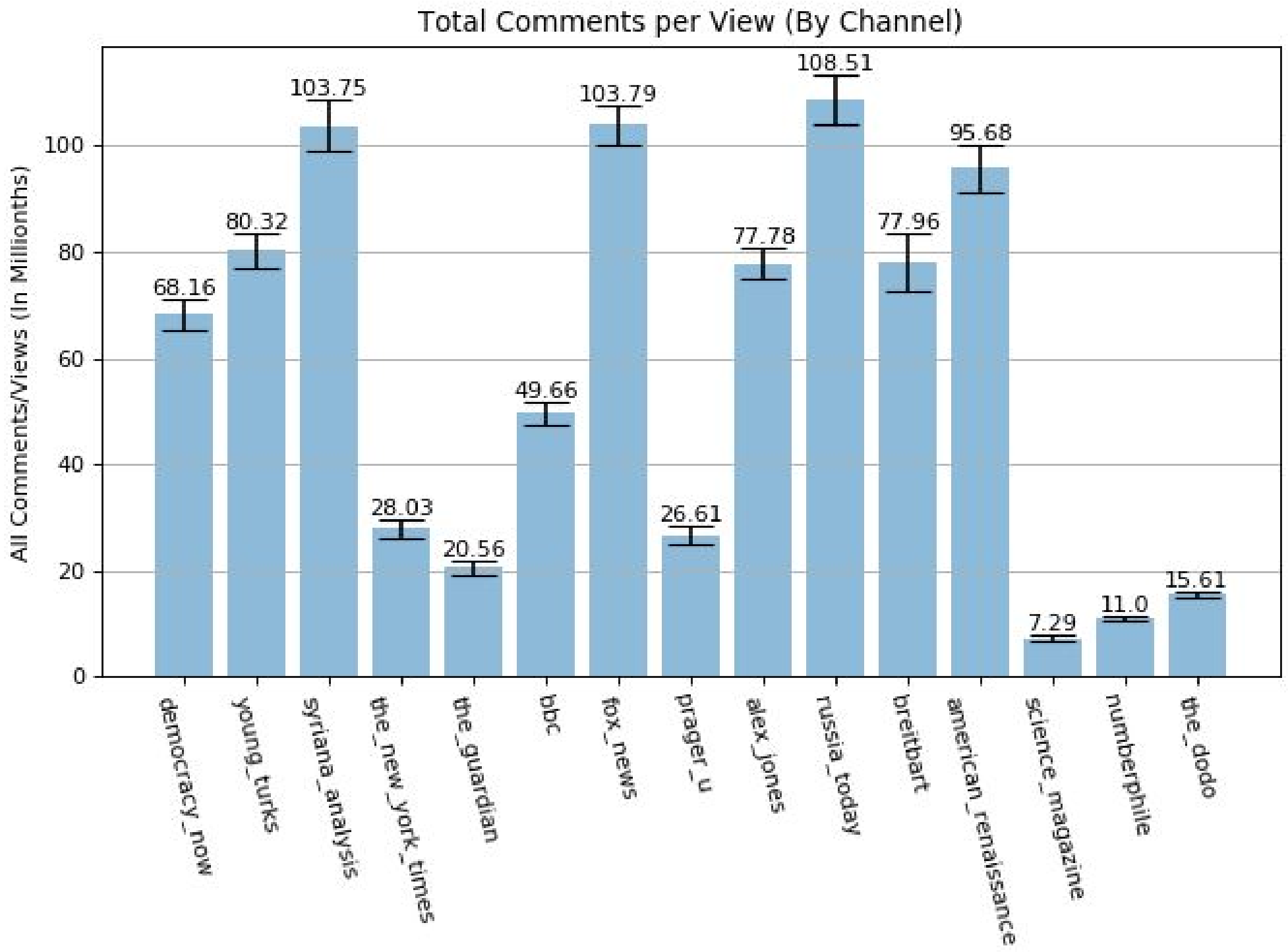

**Figure 3B: Total CPV, by category. Total view counts per channel are provided in our appendix item 3.**

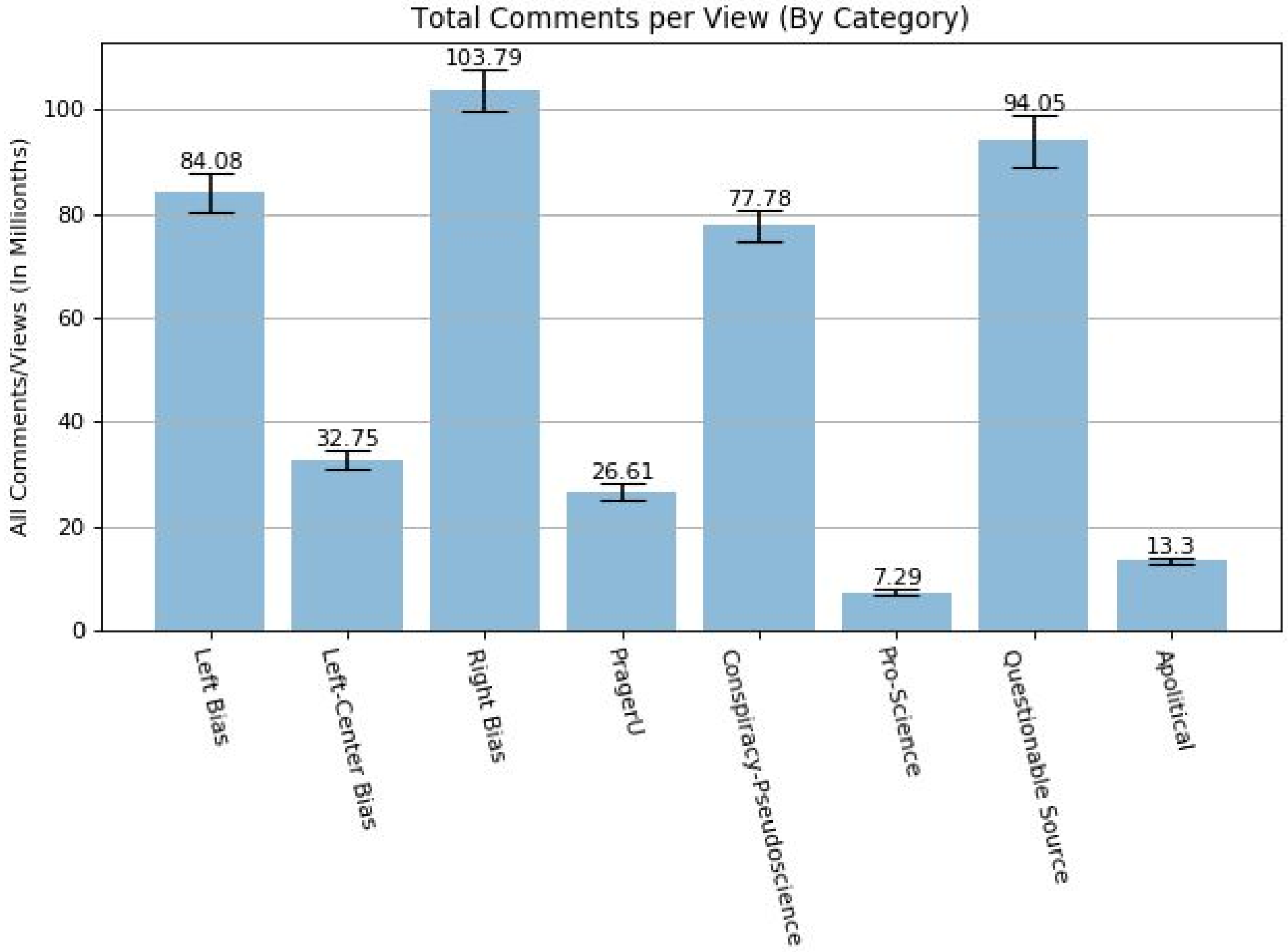

After normalizing, we find that both left-leaning and right-leaning political content resulted in more comments and replies per view as compared to non-political content. Content labelled as Left-Bias, Right-Bias, Conspiracy/Pseudoscience, or Questionable Source has 7.5x more comments per view and 10.42x more replies per view on average compared to content labelled as Apolitical or Pro-Science (Figure 3B). This indicates greater engagement in the comment sections on political channels. For our readers' convenience, we also include a finer-grain breakdown of comments per view per channel (Figure 3A).

Our data also suggests that more polarized content receives greater engagement than less polarized content. Most notably, we find 8.3x more comments per view on channels labelled as Conspiracy-Pseudoscience and Questionable Sources compared to channels labelled as Apolitical or Pro-Science. We also see that Left-Bias channels have 2.56x more comments per view than Left-Center channels on average.

Although we saw the difference between Conspiracy-Pseudoscience/ Questionable Sources and Apolitical/ Pro-Science, that does not take into account the great difference in views between the four categories; there are far more views on our Apolitical channels than on channels in the other three categories (Figure 2).

*Comments per video*

Focusing on comments per view can sometimes downplay engagement on more popular channels, since channels garnering high viewership do not always receive a proportional number of direct comments. Thus, we also wanted to make a baseline measurement by tracking the average responses per video, as opposed to per view, by channel (Figure 4A) and by category (Figure 4B).

**Figure 4A: Total comments, including both direct comments and replies, divided by videos per channel.**

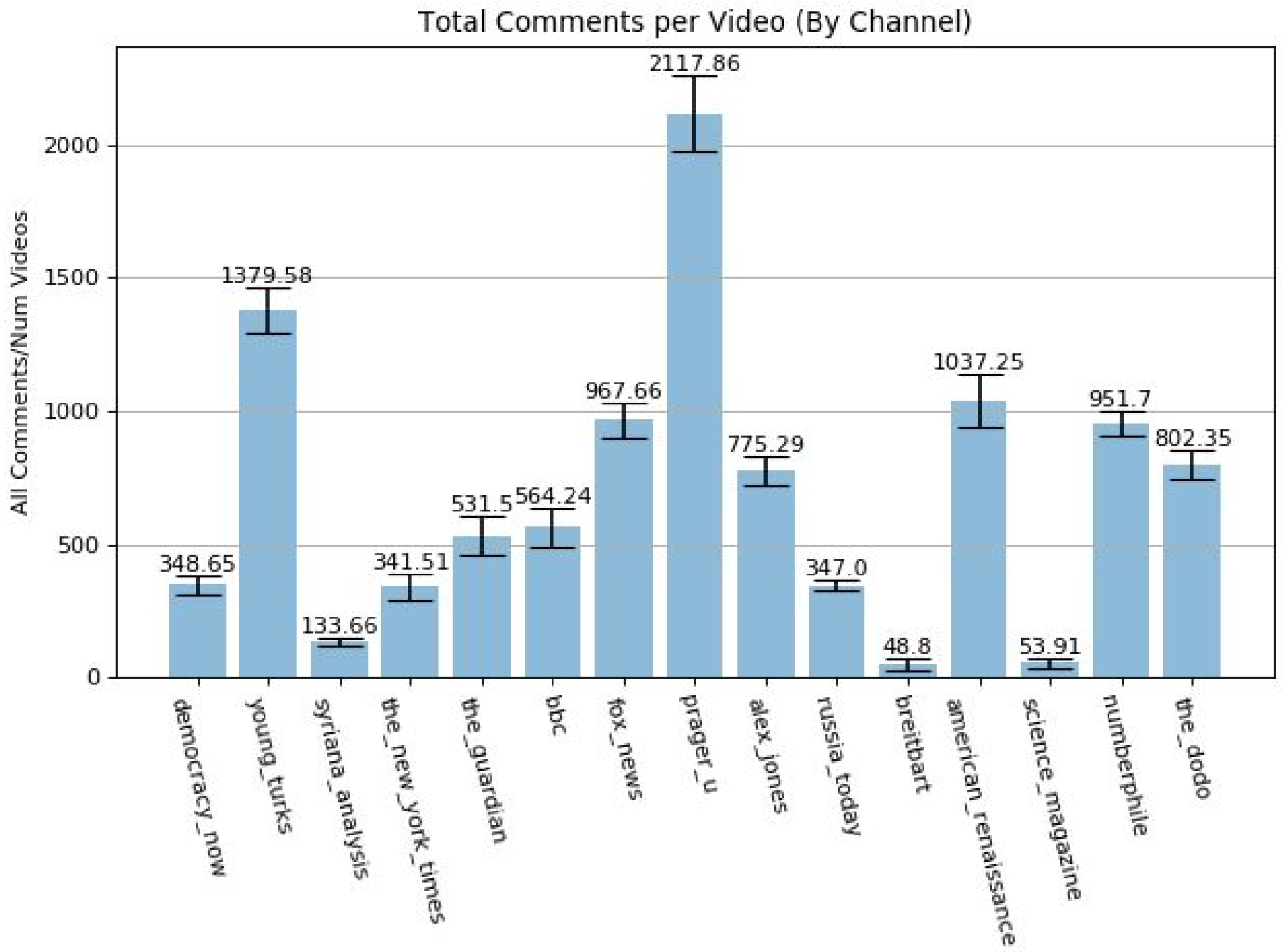

**Figure 4B: Total comments divided by videos per channel, averaged over category.**

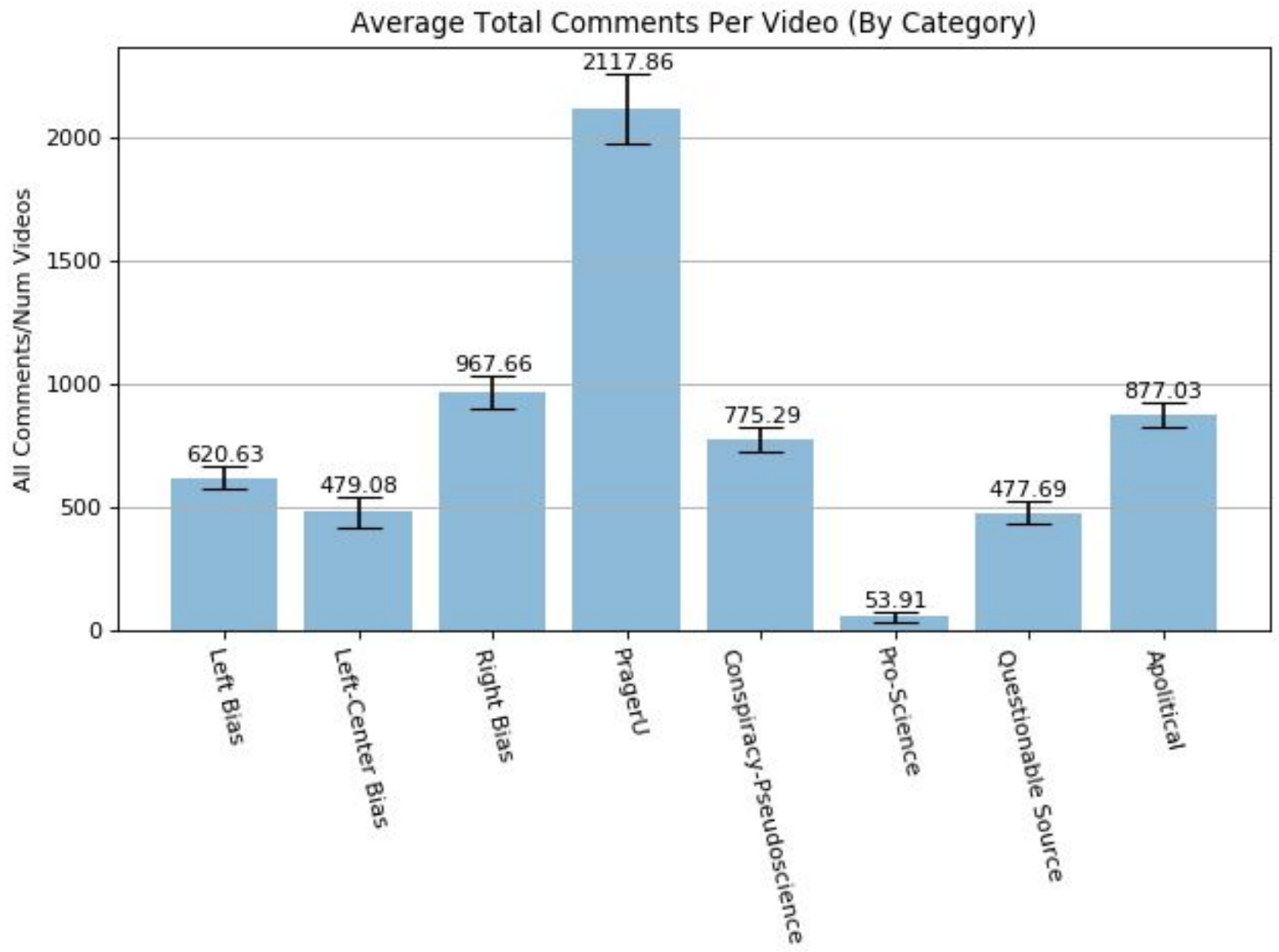

*Average thread length (ATL)*

Since we are also interested in the depth of interaction between users in YouTube comment sections, we also measure the **average thread length (ATL)** across each channel (Figure 5A) and category (Figure 5B). We consider ATL a good measure of deeper engagement because they require viewers to interact with other commenters over longer periods of time. We calculate ATL's for channels by taking the average number of replies on each direct comment in that channel, then for category by averaging the ATL's of the channels in that category.

**Figure 5A: Average thread length for each channel. Average thread length is the average number of replies on each direct comment over all the videos we scraped for a particular channel.**

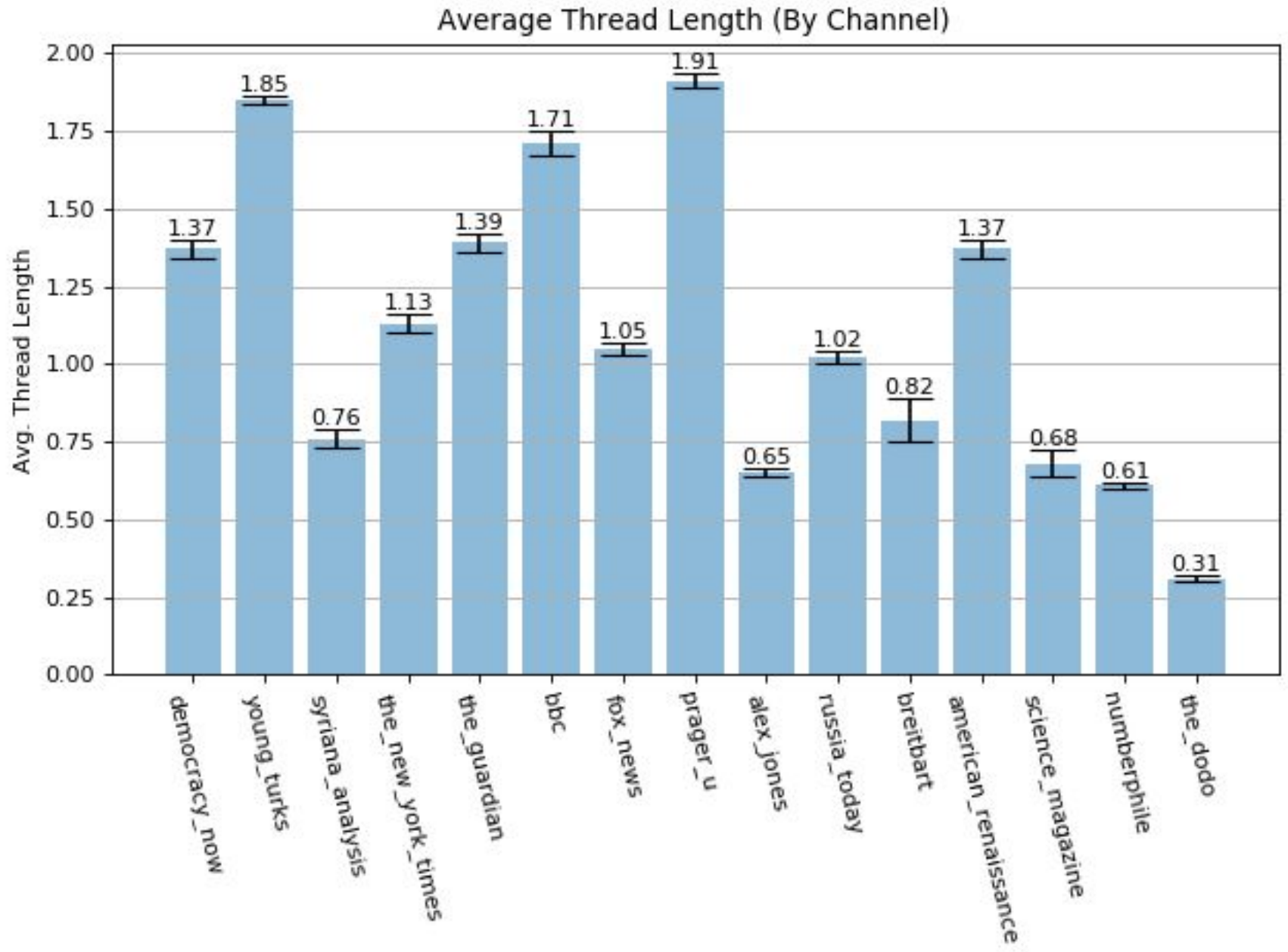

**Figure 5B: Average thread length organized by category.**

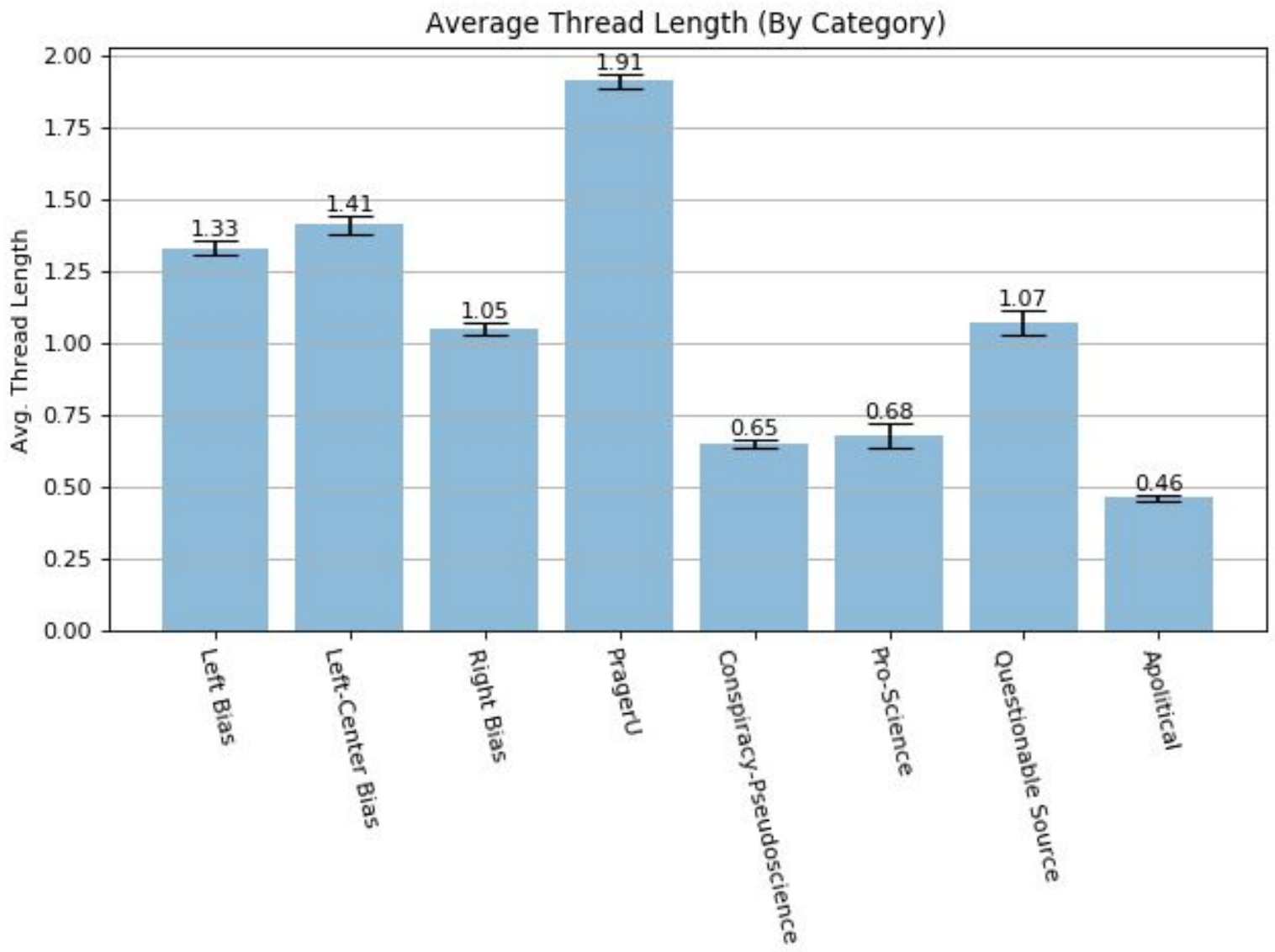

Despite having higher total view counts per video than most of the news sources (Figure 2), the apolitical channels The Dodo and Numberphile both have significantly shorter ATL's than other news sources. All our apolitical channels have ATL's < 1. Shorter ATL's is not unique to Apolitical channels; we do see that some Left-Bias, Questionable Source, Pro-Science and Conspiracy-Pseudoscience channels have ATL's < 1 (Figure 5A). However, those particular channels do not have view counts per video which are comparable to those of the Apolitical channels. If we compare the Apolitical ATL (with an average of 400k views per video) to the PragerU ATL (with 1,050k views per video), we see that PragerU's ATL is 1.5 comments more than the Apolitical ATL.

For our chosen channels, the Apolitical ATL is much less than the ATL of any other category. This broadly hints that political content leads to greater engagement.

Some further work on ATL can be found in our appendix, items 5 & 6.

*Average comment lengths (ACL)*

As another measure of comment engagement, we analyze the **average comment lengths (ACL's)** for each channel. ACL by channel takes the average of the number of characters per comment for every video in the channel's dataset (Figure 6A). ACL by category is taken by taking the averages of the ACL's of the channels in each category (Figure 6B).

**Figure 6A: Average number of characters in a comment, organized by channel.**

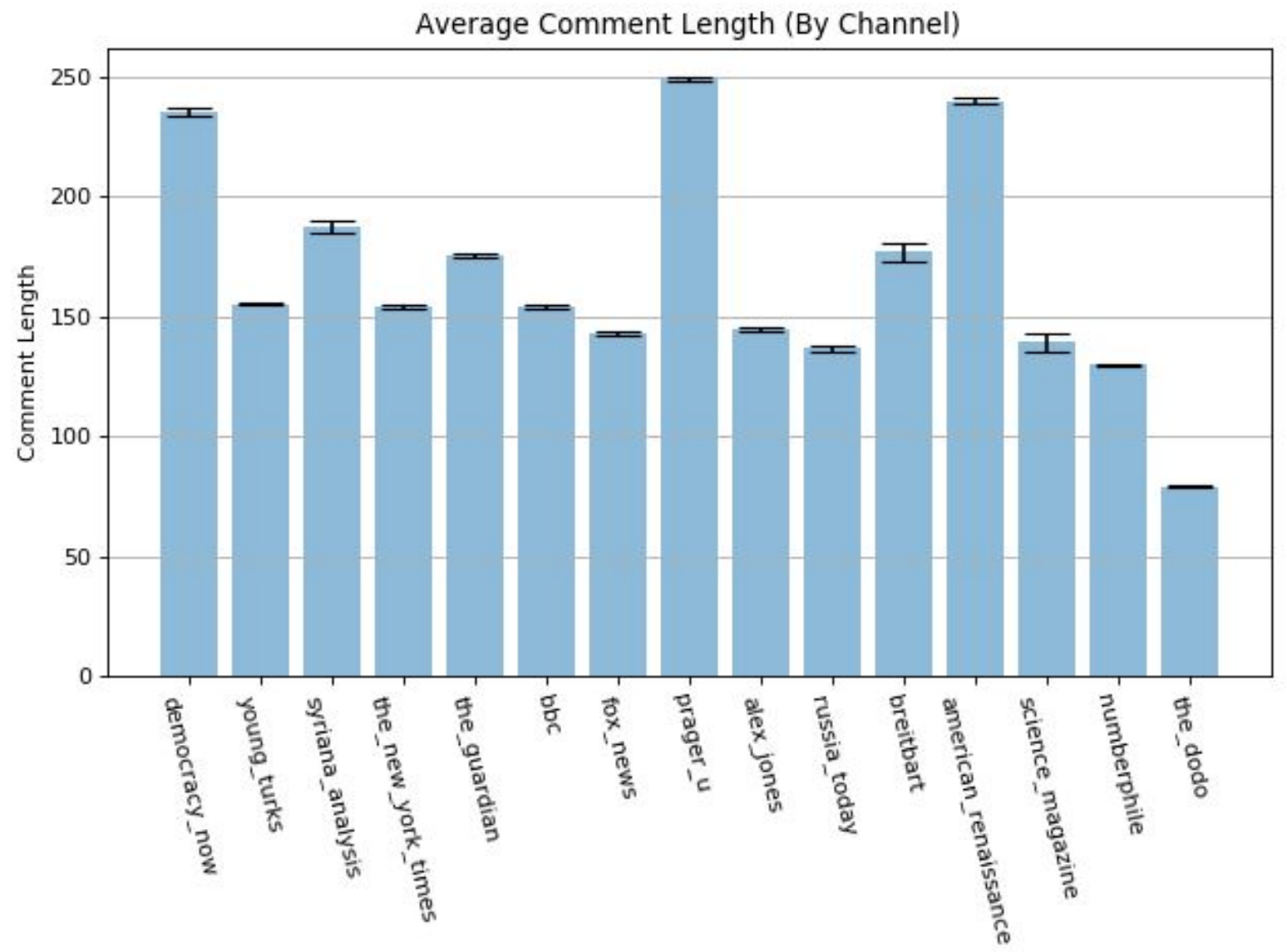

**Figure 6B: Average comment length, organized by category.**

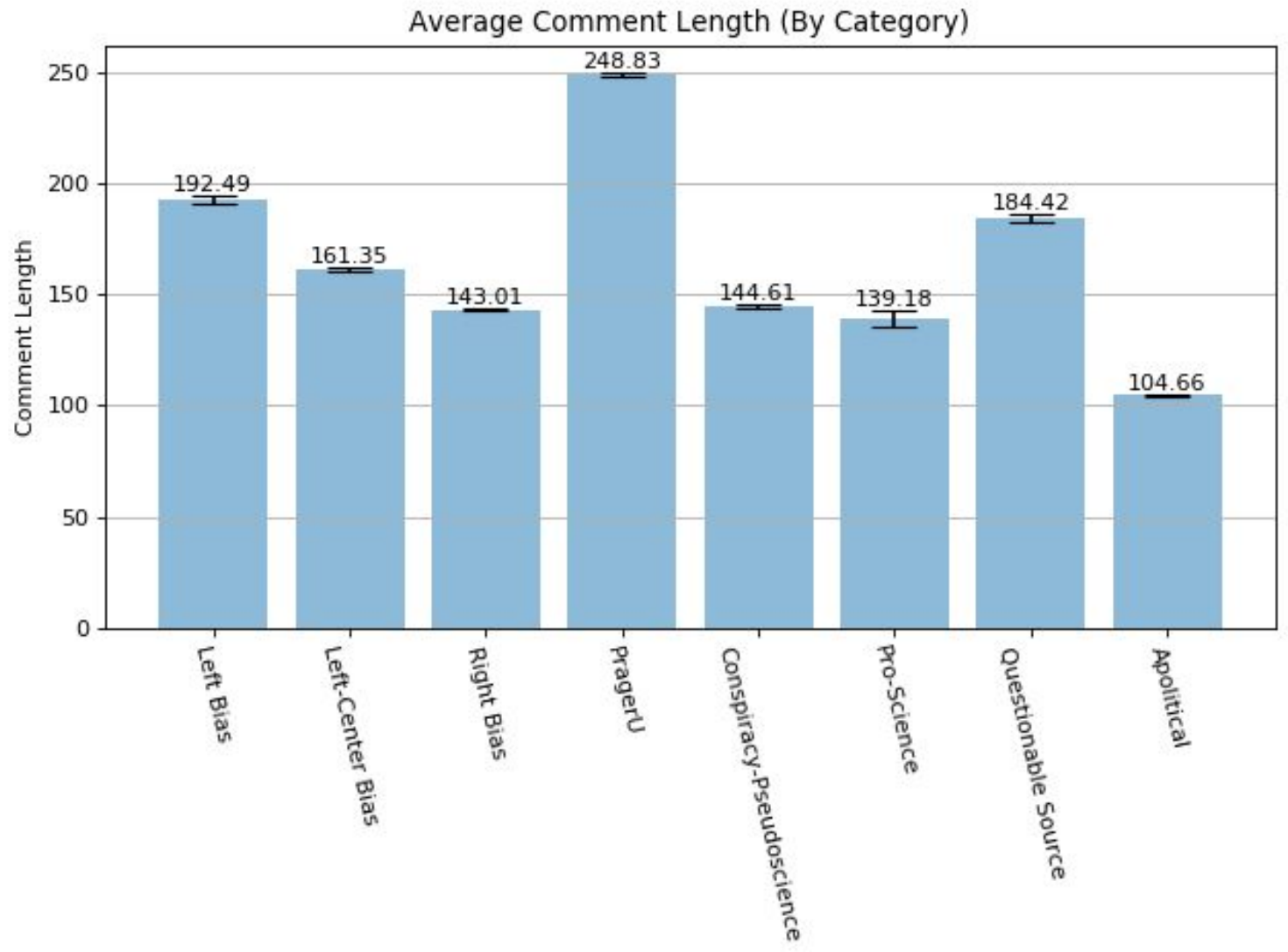

Immediately, we see that comments on PragerU were longer than comments on any other channel (Figure 6A); people are not only creating long threads (Figure 5A), but also using more words in each comment. We see similar behavior across the Democracy Now and American Renaissance channels, suggesting there may be longer comments on videos reflecting more polarized political views (Figure 6A). The Dodo lags behind.

*Profanity trends on political content*

We wanted to see if political channels inspired more profanity than apolitical channels, and whether or not there would be a relationship between MBFC ratings and profanity.

We calculate **channel profanity** by finding the percentage of profane comments per video, then averaging those percentages over all the videos in a channel (Fig. 7A); **category profanity** is calculated similarly but averaged over all videos in an MBFC category (Fig. 7B). **Profane comments** contain at least one profane word. Profanityfilter, a universal Python library for detecting profane words, checks each comment against a dictionary of profane words.

**Figure 7A: Channel profanities for our dataset.**

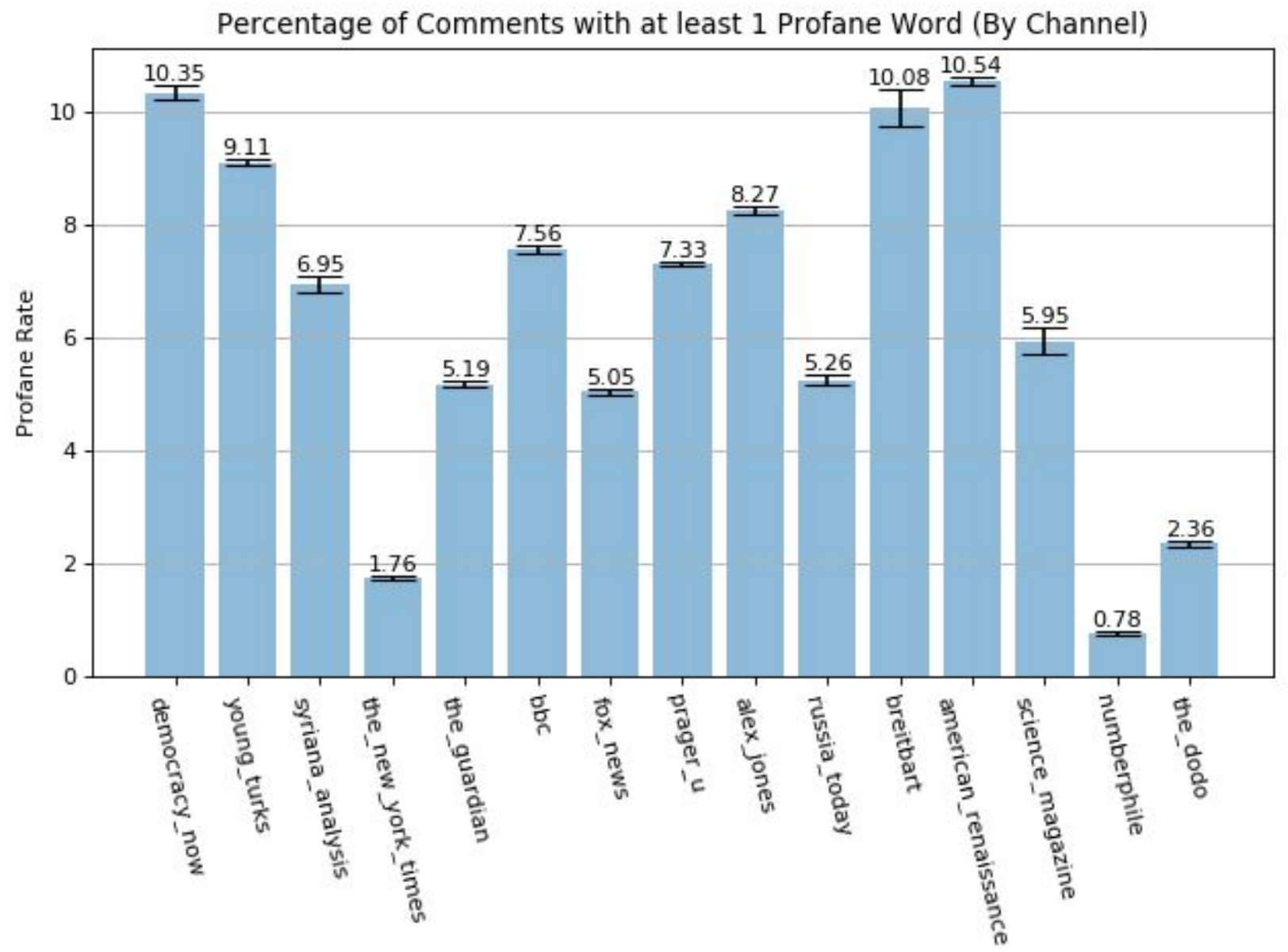

**Figure 7B: Category profanities for our dataset.**

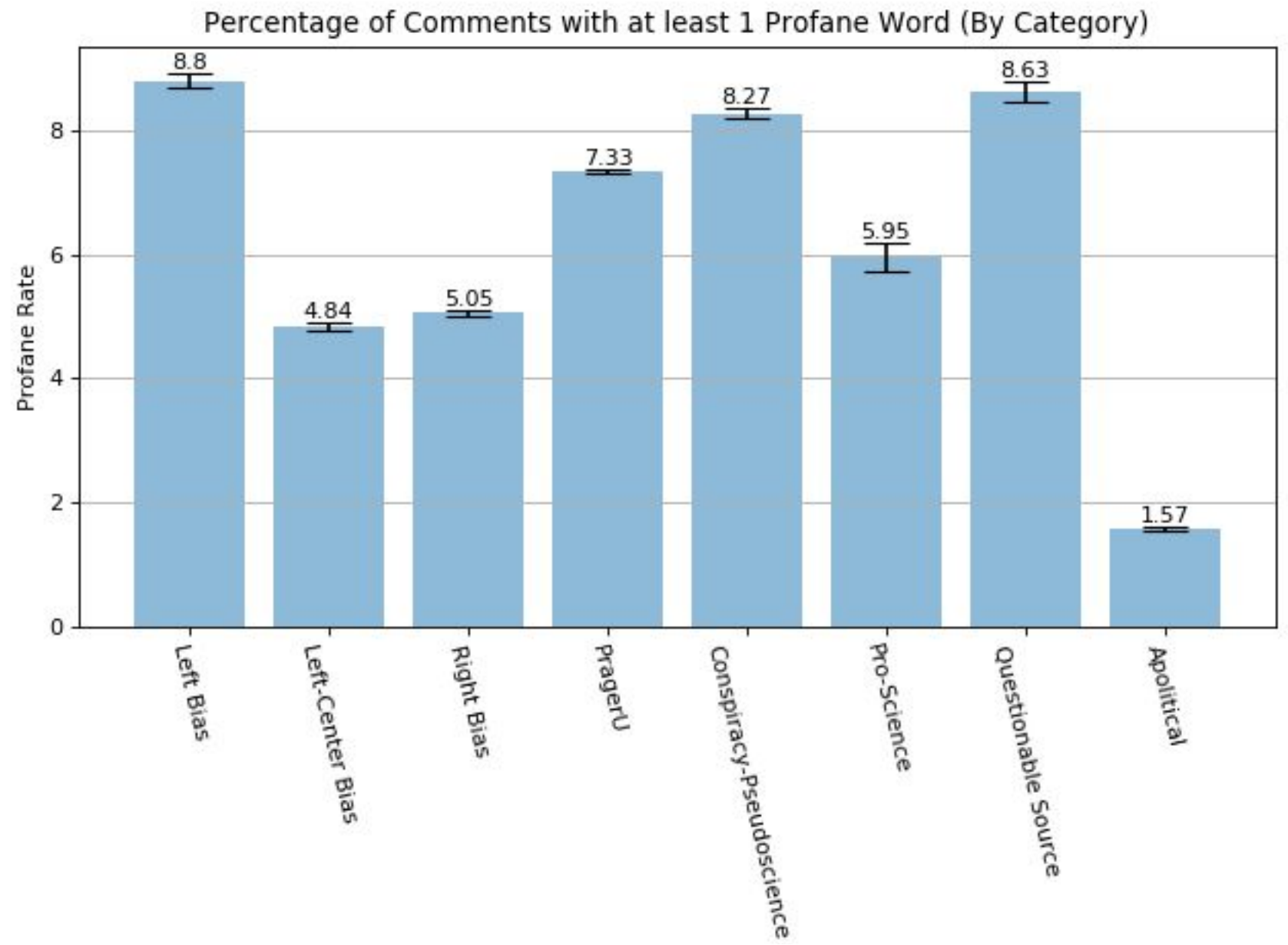

Channels with political content have 1.82x more profane comments across their videos when compared to our apolitical and pro-science sources. We also see that Left-Bias channels are 1.82x more profane than Left-Center Bias channels.

We note that Science Magazine (Pro-Science, High Factual Rating), BBC and The Guardian (both Left-Center Bias, High Factual Rating) have comparable levels of profanity compared to Russia Today (Questionable Source), Fox News (Right-Bias, Mixed Factual Rating), and Syriana Analysis (Left-Bias, Mixed Factual Rating), which suggests that factual rating is not necessarily immediately correlated with profanity. Even so, it is interesting to note that the most profane channels, American Renaissance (Questionable Source), Democracy Now (Left-Bias, Mixed Factual Rating), Young Turks (Left-Bias, Mixed Factual Rating), and Breitbart (Questionable Source), had some of the lowest factual ratings.

We made a number of word clouds per channel in the hopes of seeing what commenters on each channel were saying, but drew no conclusions. The word clouds can be found in our appendix (item 4).

**Predicting the MBFC Category of a Video using Metadata**

After observing notable differences in engagement statistics from channel to channel, we were interested in seeing how well we could leverage these differences for prediction. We selected four basic machine learning models from the scikit-learn package for Machine Learning in Python (Decision Tree, Random Forest, Linear SVM, and SVM with RBF kernel), and trained these models to predict the bias category which a given video comes from.

*Summary*

For our limited dataset, the Random Forest and SVM with RBF kernel are the highest performing models, achieving test accuracies of 80.2% and 79.0% respectively in this 8-class classification problem. The most important features used by the Decision Tree and Random Forest models to determine the bias category of a video were "Like Count" of the video, "Comments Per View", and "Profanity Rate". In future work, we hope to increase the size of our dataset and leverage NLP techniques on the text of comments to improve classification performance and gain more insight on differences in engagement and language between videos from different political bias categories.

*Predicting the MBFC category of a video*

We combined the videos of our 15 YouTube channels into one large dataset of 2904 videos. Each video was given 10 features, shown in Figure 8A below. These features consist of those engagement statistics which we compared earlier (ex. "Profanity Rate") and other metadata which we had easy access to (ex. "Like Count").

Figure 8A depicts the importance of these features according to our Decision Tree and Random Forest models, where **feature importance** is defined as the normalized total reduction of the split criterion (Gini impurity) provided by that feature.

It is worth noting the slight redundancy in using "Comments Per View", "Number of Comments", and "Number of Views" as features; the first is the ratio of the other two. Interestingly, the Decision Tree and Random Forest models find that the "Comments Per View" is a substantially more informative feature than the two statistics alone.

**Figure 8A: A list of the features we used, as well as each one's importance in Decision Tree and RF**

| Feature | Decision Tree Importance | Random Forest Importance |
|---:|---:|---:|
| Num Comments | 0.028195 | 0.067250 |
| Average Direct Comment Like Count | 0.071287 | 0.073852 |
| Num Unique Authors | 0.029831 | 0.076802 |
| Average Thread Length | 0.066847 | 0.079858 |
| Average Comment Length | 0.084224 | 0.085174 |
| Dislike Count | 0.124162 | 0.098351 |
| Num Views | 0.079629 | 0.107514 |
| Profanity Rate | 0.158240 | 0.121183 |
| Comments Per View | 0.133817 | 0.136023 |
| Like Count | 0.223767 | 0.153991 |

We visualized the 10-dimensional data, labelled by its bias category, using 2-dimensional t-distributed Stochastic Neighbor Embedding in Figure 8B. The most visible trait of this visualization is that Apolitical videos, which had the lowest average value for many of the features in the original feature space (e.g. “Comments Per View”, “Average Thread Length”, “Profanity Rate”), form a relatively independent cluster.

**Figure 8B: A 2-dimensional t-SNE visualization of the training data.**

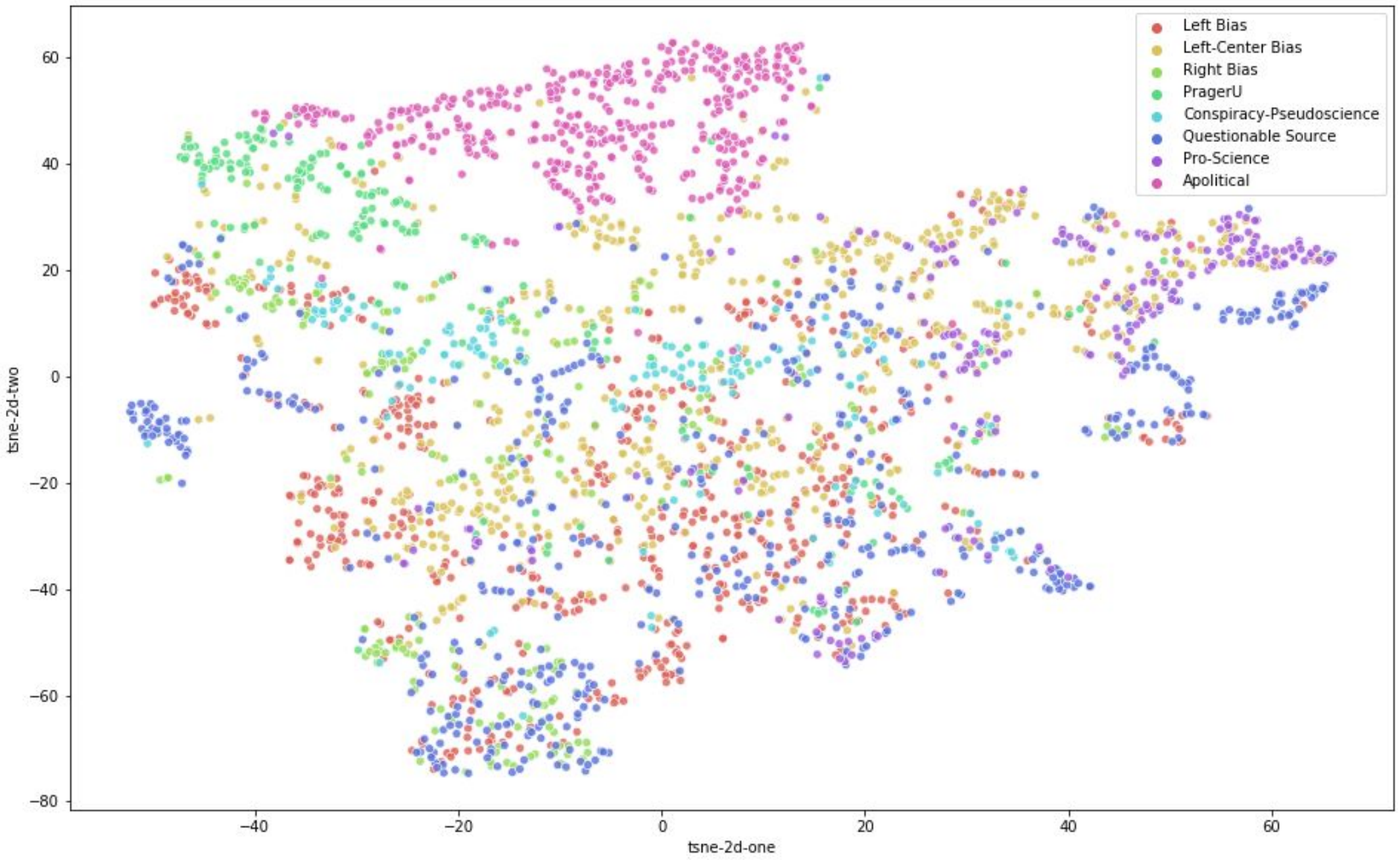

For each model applied to the data, we used 5-fold cross validation on our training set (85% of the total dataset) to find optimal hyper-parameters. After training our models on those hyperparameters, we evaluated them on an unseen test set (15% of the total dataset). The resulting confusion matrices on the unseen test data for each model can be found in Figure 8C.

We find that the Random Forest and SVM with RBF kernel perform best, with test accuracies of 80.2% and 79.0% respectively.

We know that Random Forest models generally perform well on any dataset with minimal regularity because of their ensemble nature and their robustness to outliers. It is no surprise that the linear SVM underperforms because there is no reason to believe that the optimal decision function for this problem would be a linear decision boundary. Thus, we hypothesize that the linear SVM suffers from bias to a greater extent than the other models.

The SVM with RBF kernel likely performs well because the RBF kernel function can be interpreted as a similarity score based on Euclidean distance between points in the feature space. Videos from the same channel tend to have similar feature representations, and each bias category in our dataset is composed of videos from no more than three unique channels.

**Figure 8C: Confusion matrices for models with hyper-parameters chosen by 5-fold cross validation. (Top left) Decision tree with maximum depth = 12: test accuracy = 65.7%. (Top right) Random forest with unlimited max depth, 61 estimators: test accuracy = 80.2%. (Bottom left) SVM with C = 100: test accuracy = 62.5%. (Bottom right) SVM with RBF kernel, with gamma = 0.1, C = 175: test accuracy = 79.0%.**

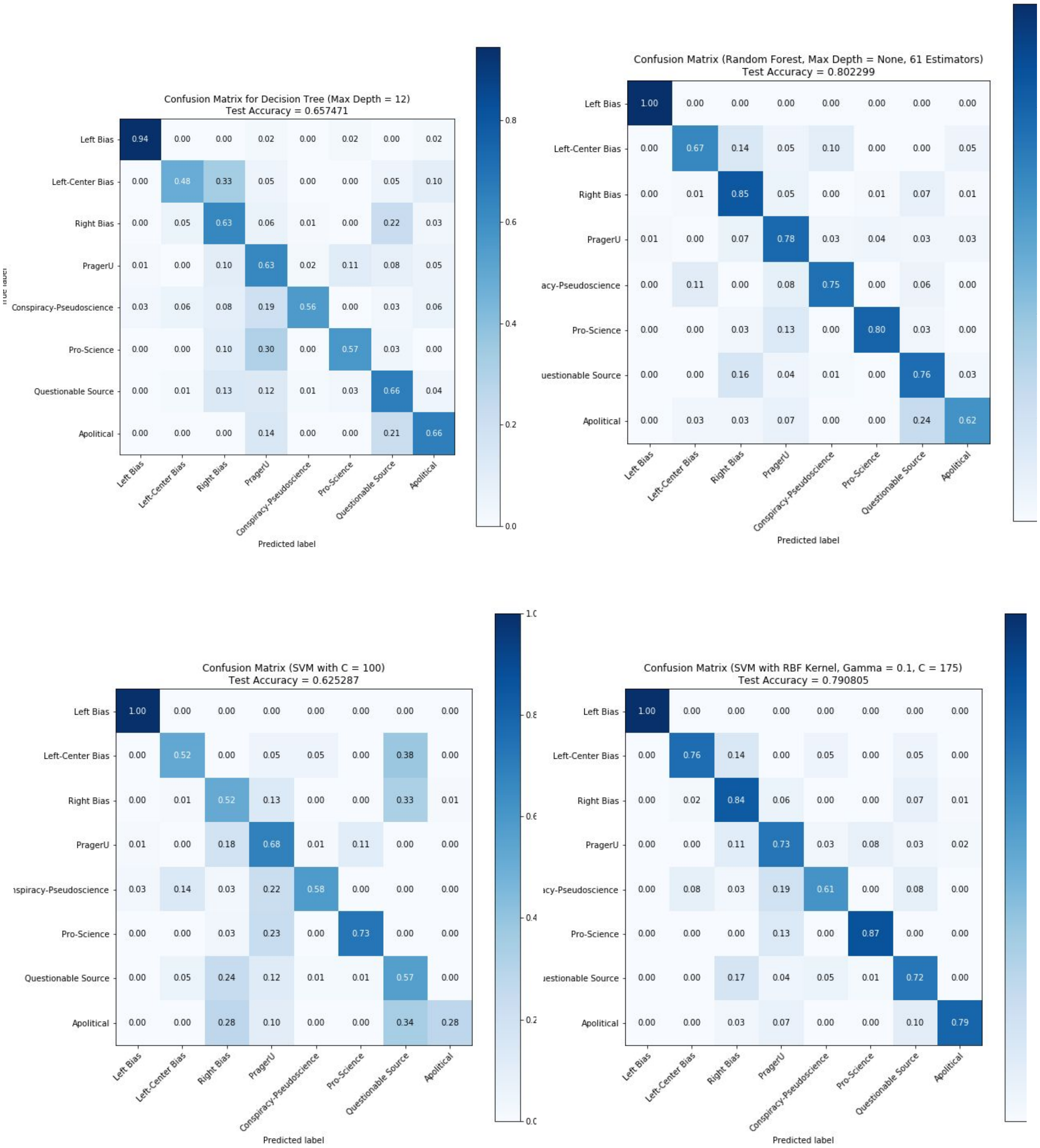

For future work, it would be interesting to see if the SVM with RBF kernel continues performing well with greater amounts of broader data. We are also interested in applying NLP techniques to the comments of videos to gain novel insights about differences in the language of comments in videos from different bias categories.

## Discussion

Our preliminary findings indicate notably higher viewer engagement on biased, political, or misinformative channels when compared to unbiased, apolitical, and truthful channels, appearing to corroborate similar research in this field. Using Decision Tree and Random Forest classifiers on our dataset suggested that “Like Count”, “Comments per View”, and “Profanity Rate” were the most important features for determining the bias of a channel or whether it was classified as “Conspiracy-Pseudoscience” or a “Questionable Source”.

Having limited our scope to 15 YouTube channels, a majority of which were American, it remains to be seen if the trends discussed in this paper occur across a larger and more comprehensive dataset. Furthermore, the differences in viewership across the chosen channels may have led us to different conclusions than if the compared channels had more similar viewership. These uncertainties could be resolved by running the same analyses on a larger, more diverse dataset. We would also like to expand our scope to analyze why certain comments receive greater traction and why some channels have longer average thread lengths, as well as how discussions and arguments develop on YouTube.

We welcome fellow researchers and the open-source community to use and build upon our work. We hope our readers remember that online comments and discussions may never accurately reflect the nature of people’s attitudes towards the content they see - many interactions happen offline, “in real life” - and look to stimulate further analysis in the domain of discourse surrounding misinformation online.

# Appendix

## Item 1: MBFC Categories and Factual Ratings

These definitions are sourced directly from Media Bias Fact Check. Their methodology can be found at https://mediabiasfactcheck.com/methodology/. We have only provided descriptions for the categories and factual ratings for our channels.

*Category details*

- **Left Bias**: "These media sources are moderately to strongly biased toward liberal causes through story selection and/or political affiliation. They may utilize strong loaded words (wording that attempts to influence an audience by using appeal to emotion or stereotypes), publish misleading reports and omit reporting of information that may damage liberal causes. Some sources in this category may be untrustworthy."
- **Left-Center:** "These media sources have a slight to moderate liberal bias. They often publish factual information that utilizes loaded words (wording that attempts to influence an audience by using appeal to emotion or stereotypes) to favor liberal causes. These sources are generally trustworthy for information, but may require further investigation."
- **Right Bias:** "These media sources are moderately to strongly biased toward conservative causes through story selection and/or political affiliation. They may utilize strong loaded words (wording that attempts to influence an audience by using appeal to emotion or stereotypes), publish misleading reports and omit reporting of information that may damage conservative causes. Some sources in this category may be untrustworthy."
- **Pro-Science:** "These sources consist of legitimate science or are evidence based through the use of credible scientific sourcing. Legitimate science follows the scientific method, is unbiased and does not use emotional words. These sources also respect the consensus of experts in the given scientific field and strive to publish peer reviewed science. Some sources in this category may have a slight political bias, but adhere to scientific principles."
- **Conspiracy-Pseudoscience:** "Sources in the Conspiracy-Pseudoscience category may publish unverifiable information that is not always supported by evidence. These sources may be untrustworthy for credible/verifiable information, therefore fact checking and further investigation is recommended on a per article basis when obtaining information from these sources."
- **Questionable Sources:** "A questionable source exhibits one or more of the following: extreme bias, consistent promotion of propaganda/conspiracies, poor or no sourcing to credible information, a complete lack of transparency and/or is fake news. Fake News is the deliberate attempt to publish hoaxes and/or disinformation for the purpose of profit or influence (Learn More). Sources listed in the Questionable Category may be very untrustworthy and should be fact checked on a per article basis. Please note sources on this list are not considered fake news unless specifically written in the reasoning section for that source."

*Factual rating details:*

- **Very Low**: "the source almost never uses credible sources and is simply not trustworthy for reliable information at all. These are the sources that always need to be fact checked for intentional fake news, conspiracy and propaganda."
- **Mixed**: "the source does not always use proper sourcing or sources to other biased/mixed sources. They may also report well sourced information as well. Mixed sources will have failed one or more fact checks and does not immediately correct false or misleading information. Further, any source that does not disclose a mission statement or ownership information will automatically be deemed Mixed as will a source that utilizes extremely loaded language that alters context of facts, but is still properly sourced and has not failed a fact check. Lastly, any source that does not support the consensus of science on such topics as Climate Change, GMO, Vaccinations, Evolution, or any other will automatically be rated Mixed for factual reporting."
- **High**: "the source is almost always factual, sources to mostly credible low biased or high factual information and makes immediate corrections to incorrect information, has failed only 1 fact check and uses reasonable language that retains context."
- **Very High:** "the source is always factual, sources to credible information and makes immediate corrections to incorrect information and has never failed a fact check."

---

## Item 2: Channel Details

For each of the 15 channels whose videos we've scraped, we provide the date when we scraped the channels' videos, basic YouTube statistics, categorizations/factual ratings (when available), and descriptions for each channel. Of these, we have released 14 datasets.

*Details of the descriptions:*

1. Dates of scraping are our own metadata. Our dates of scraping vary from December 2018 to February 2019.
2. YouTube channel metadata was gathered on December 7, 2018, for all channels except for American Renaissance, which we checked on February 27, 2019.
3. Categorizations/factual ratings were taken from MBFC on February 28, 2019 for fourteen of our seventeen sources. For The Dodo, Numberphile, and PragerU, we could not find MBFC's ratings, but proceeded to label them unofficially as apolitical, apolitical, and political (right-bias) for reference through our paper.
4. Descriptions of each channel were sourced from their own website if we could find it, their YouTube channel if there was no description on their website, and from Wikipedia if we could not find a description on either of those. Newline characters have been removed from the descriptions for formatting purposes.

*Details*

1. Democracy Now
    a. Scraped on: December 2, 2018
    b. 444,805 subscribers, 9,578 videos, joined Jul 12, 2006, 124,722,701 views
    c. MBFC: full report at https://mediabiasfactcheck.com/democracy-now/
        i. Factual Reporting: “High”
        ii. Category: “Left Bias”
    d. From https://www.democracynow.org/about: Democracy Now! produces a daily, global, independent news hour hosted by award-winning journalists Amy Goodman and Juan González. Our reporting includes breaking daily news headlines and in-depth interviews with people on the front lines of the world’s most pressing issues. On Democracy Now!, you’ll hear a diversity of voices speaking for themselves, providing a unique and sometimes provocative perspective on global events.
2. Young Turks
    a. Scraped on: December 1, 2018
    b. MBFC: full report at https://mediabiasfactcheck.com/the-young-turks/
        i. Factual Reporting: “Mixed”
        ii. Category: “Left Bias”
    c. 4,224,662 subscribers, 34,938 videos, joined Dec 21, 2005, 4,515,291,409 views
    d. From https://legacy.tyt.com/about/ : TYT Network is the largest online talk, news and entertainment networks for the connected generation. The award-winning TYT is one of the top multi-platform online content creators, generating over 200 million views a month. According to the most recent comScore ratings, TYT ranks #1 in News and Politics on all digital platforms among the millennial audience (18-24).
3. Syriana Analysis
    a. Scraped on: December 3, 2018
    b. 27,432 subscribers, 303 videos, joined Jan 13, 2017, 5,698,274 views
    c. MBFC: full report at https://mediabiasfactcheck.com/syriana-analysis/
        i. Factual Reporting: “Mixed”
        ii. Category: “Left Bias”
    d. From https://www.syriana-analysis.com/about/ : Syriana Analysis is dedicated to covering the current affairs in Syria. Launched in January 2017, Syriana Analysis is an independent Syrian news and analysis platform free from any political or corporate funding, sponsorship or association. Syriana Analysis is exclusively funded by its subscribers on YouTube, which gives us the privilege to report the events from our perspective and what makes us believe serving the best

interests of most Syrians. Syriana Analysis is pro-united and secular Syria and anti-religious/ethnic fundamentalism and separatism.

4. New York Times
    a. Scraped on: December 4, 2018
    b. 1,610,431 subscribers, 9,415 videos, joined Oct 13, 2006, 669,650,991 views
    c. MBFC: full report at https://mediabiasfactcheck.com/new-york-times/
        i. Factual Reporting: "High"
        ii. Category: "Left-Center Bias"
    d. From https://www.youtube.com/user/TheNewYorkTimes/about: The New York Time is the most powerful engine for independent, boots-on-the-ground and deeply reported journalism. We set the standard for the most ambitious and innovative storytelling across features, news and investigations. Because we're journalists, we're excited to report the news as quickly as possible, use new technological resources to uncover the truth, and unearth new cultural phenomenons with our critics. The internet didn't plant these ideas in our heads. We've always been this way. It's all the news that's fit to watch. On YouTube.
5. BBC
    a. Scraped on: December 1, 2018
    b. 3,483,096 subscribers, 9,282 videos, joined Apr 7, 2006, 934,374,518 views
    c. MBFC: full report at https://mediabiasfactcheck.com/bbc/
        i. Factual Reporting: "High"
        ii. Category: "Left-Center Bias"
    d. From https://www.youtube.com/user/bbcnews/about : Welcome to the official BBC News YouTube channel. Interested in global news with an impartial perspective? Want to see behind-the-scenes clips and footage directly from the front-line? Our YouTube channel has all this and more, bringing you specially selected clips from the world's most trusted news source. Tune into BBC World News for 24 hour news on TV. Check out BBC News on mobile and download our app for iPhone, Android and Kindle Fire. The official BBC News YouTube channel is operated by BBC Global News Ltd which comprises BBC World News the 24hr TV channel and bbc.com/news, the international news website.
6. The Guardian
    a. Scraped on: December 3, 2018
    b. 724,629 subscribers, 6,518 videos, joined Feb 14, 2006, 264,488,634 views
    c. MBFC: full report at https://mediabiasfactcheck.com/the-guardian/
        i. Factual Reporting: "High"
        ii. Category: "Left-Center Bias"
    d. From https://www.youtube.com/user/TheGuardian/about: The latest news features, documentaries and opinion videos from The Guardian.

7. The Dodo
    a. Scraped on: December 1, 2018
    b. 3,046,148 subscribers, 4,744 videos, joined Mar 21, 2014, 1,223,201,042 views
    c. MBFC
        i. not found
    d. From https://www.youtube.com/user/TheDodoSite/about : The Dodo on YouTube is a place for everyone who loves animals and cares about their wellbeing. Our goal is to make caring about animals a viral cause. We want our fans to fall in love with animals, be entertained while they're doing it, and feel empowered to help animals in need.
8. Numberphile
    a. Scraped on: December 1, 2018
    b. 2,683,988 subscribers, 501 videos, joined Sep 15, 2011, 370,901,776 views
    c. MBFC
        i. not found
    d. From https://www.youtube.com/user/numberphile/about : Videos about numbers - it's that simple. Videos by Brady Haran
9. Science Magazine
    a. Scraped on: December 3, 2018
    b. 119,934 subscribers, 590 videos, joined Mar 24, 2008, 30,026,700 views
    c. MBFC: full report at https://mediabiasfactcheck.com/science-magazine/
        i. Factual Reporting: "Very High"
        ii. Category: "Pro-Science"
    d. From https://www.youtube.com/user/ScienceMag/about : The latest videos from Science magazine, the world's leading outlet for scientific news, commentary, and cutting-edge research. Learn more at www.sciencemag.org
10. PragerU
    a. Scraped on: December 3, 2018
    b. 1,879,540 subscribers, 504 videos, joined Jun 10, 2009, 564,904,591 views
    c. MBFC
        i. not found
    d. From https://www.prageru.com/about/ : We take the best ideas from the best minds and distill them into short videos. Our comprehensive digital marketing campaign promotes the ideas that have made America and the West the source of so much liberty and wealth. These values are Judeo-Christian at their core and include the concepts of freedom of speech, free markets and love for our country—The United States of America.
11. RT (Russia Today)
    a. Scraped on: December 3, 2018

b. 3,234,208 subscribers, 42,075 videos, joined Mar 28, 2007, 2,622,305,013 views
c. MBFC: full report at https://mediabiasfactcheck.com/rt-news/
    i. Factual Reporting: n/a
    ii. Category: "Questionable Source"
d. From https://www.youtube.com/user/RussiaToday/about : RT is a global news channel broadcasting from Moscow and Washington studios. With a global reach of over 700 million people, or over 25% of all cable subscribers worldwide, RT news covers the major issues of our time for viewers wishing to question more. Our team of young news professionals has made RT the first news channel to break the 1 billion YouTube views benchmark. Question more - together with RT.

12. InfoWars*
    a. * InfoWars was removed from YouTube in August 2018, so we have no statistics for it from YouTube. All of their videos were also removed from YouTube, including the ones whose comments we scraped. The videos whose comment threads we included in our InfoWars dataset were posted between October 2017 and April 2018, and were scraped in April 2018.
    b. MBFC: full report at https://mediabiasfactcheck.com/infowars-alex-jones/
        i. Factual Reporting: "Very Low"
        ii. Category: "Conspiracy-Pseudoscience"
    c. From https://en.wikipedia.org/wiki/InfoWars : InfoWars is a far-right American conspiracy theory and fake news website. It was founded in 1999, and is owned by Free Speech Systems LLC.
13. Fox
    a. Scraped on: December 16, 2018
    b. 2,186,028 subscribers, 54,441 videos, joined Sep 18, 2006, 1,338,990,079 views
    c. MBFC rating: full report at https://mediabiasfactcheck.com/fox-news/
        i. Factual Reporting: "Mixed"
        ii. Category: "Right Bias"
    d. From https://www.youtube.com/user/FoxNewsChannel/aboutv : FOX News Channel (FNC) is a 24-hour all-encompassing news service dedicated to delivering breaking news as well as political and business news. A top cable network in both total viewers and Adults 25-54, FNC has been the most-watched news channel in the country for more than 15 years and according to Public Policy Polling, is the most trusted television news source in the country. Owned by 21st Century Fox, FNC is available in more than 89 million homes and dominates the cable news landscape, routinely notching 12 of the top 15 programs in the genre.
14. Breitbart News
    a. Scraped on: December 4, 2018

b. 79,288 subscribers, 1,215 videos, joined Feb 24, 2015, 14,481,023 views
c. MBFC rating: full report at https://mediabiasfactcheck.com/breitbart/
    i. Factual Reporting: n/a
    ii. Category: “Questionable Source”
d. From https://en.wikipedia.org/wiki/Breitbart_News : Breitbart News Network is a far-right syndicated American news, opinion and commentary website founded in mid-2007 by conservative commentator Andrew Breitbart, who conceived it as "the Huffington Post of the right."

15. American Renaissance
    a. Scraped on: February 27, 2019
    b. 102,486 subscribers, 175 videos, joined Oct 18, 2011, 10,957,619 views
    c. MBFC rating: full report at https://mediabiasfactcheck.com/american-renaissance-magazine/
        i. Factual Reporting: n/a
        ii. Category: “Questionable Source”
    d. From https://www.amren.com/about/ : American Renaissance was published as a monthly print magazine from October 1990 through January 2012. All back issues are available here. AR has had a web presence since 1994, and we consider AmRen.com to be the Internet’s premier race-realist site. Every weekday we publish articles and news items from a world-wide race-realist perspective.

---

### Item 3: Total views across all sampled videos for each YouTube channel. Views measured in tens of millions.

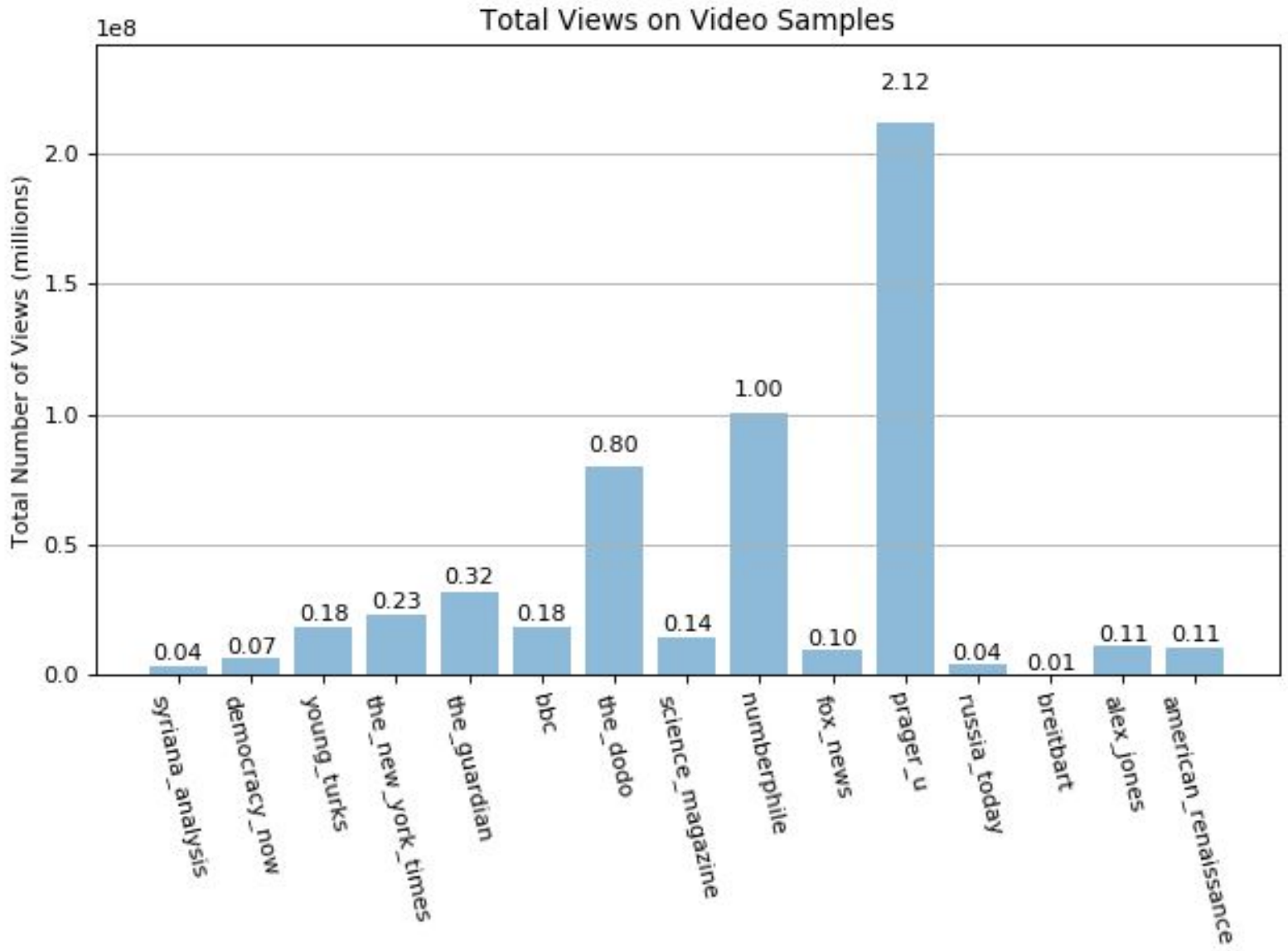

We notice that PragerU has significantly more views than the other channels. One possible reason for this difference could be that “traditional” news sources may be showing their videos on multiple different platforms, and those may not count towards view counts, while sites such as PragerU may be relying more on a singular medium (YouTube) to distribute their videos than other sites.

**Note**: this graph was made on a slightly older different version of our datasets, where each of our datasets (except for Fox, Alex Jones/Infowars, and American Renaissance) had 1 extra video’s metadata, but were otherwise the same as they are now. This discrepancy does not show in the main body of our work.

---

## Item 4: Word Cloud Visualizations

We sought to characterize the vocabulary used by different channels through word clouds. For each of the channels selected, we generate a word cloud containing the top words used in all direct comments across the channel. We also generate a single word cloud containing all the comments to get a rough baseline of shared words.

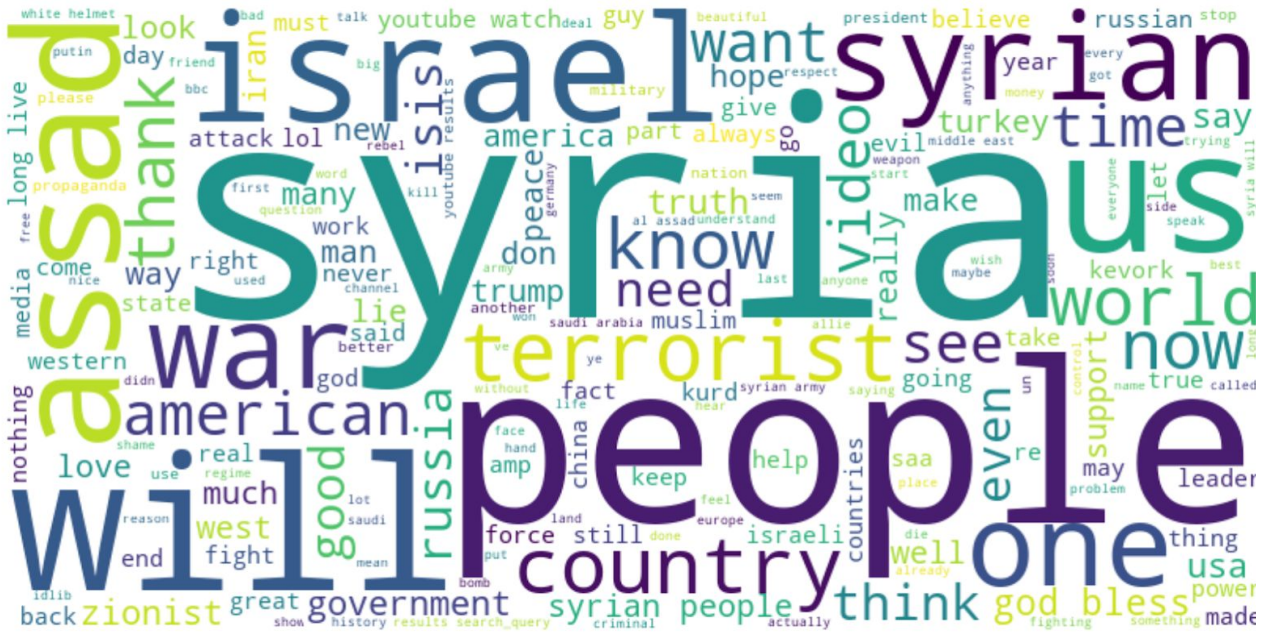

**Syriana Analysis**

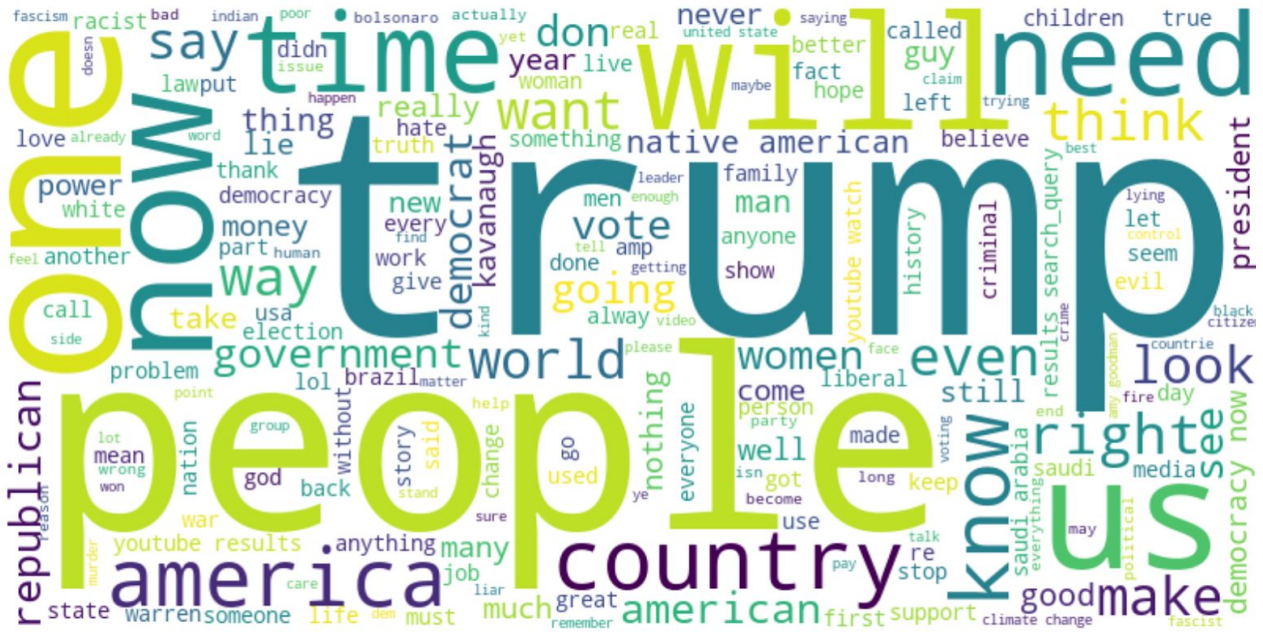

**Democracy Now**

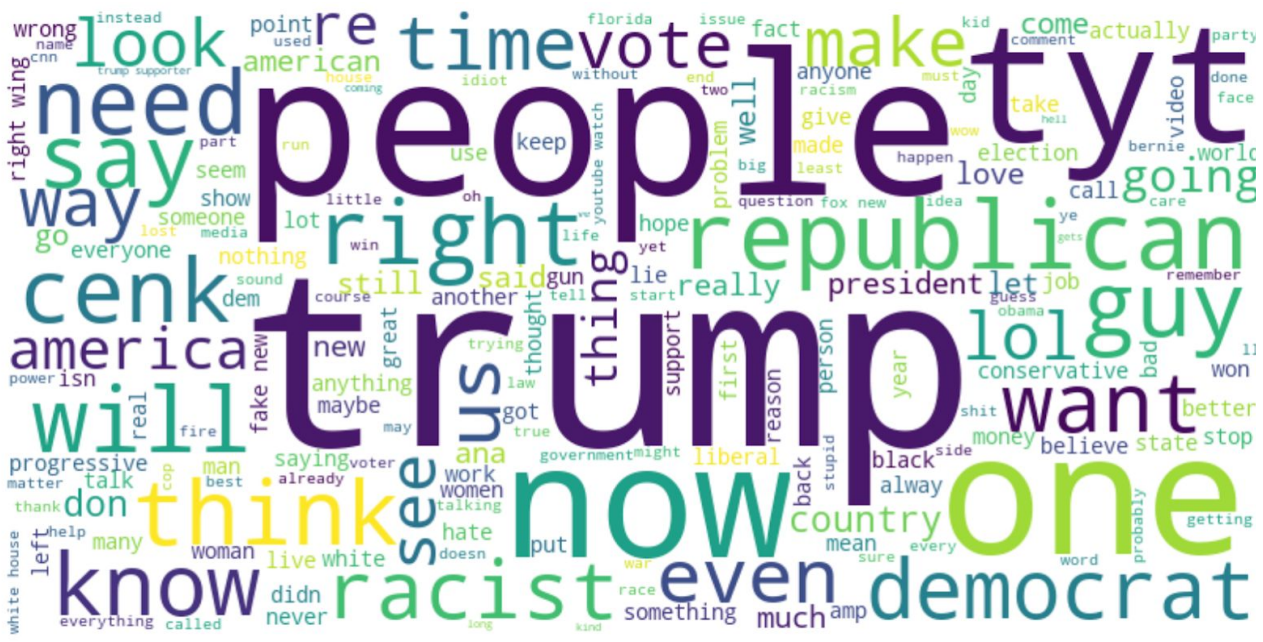

**Young Turks**

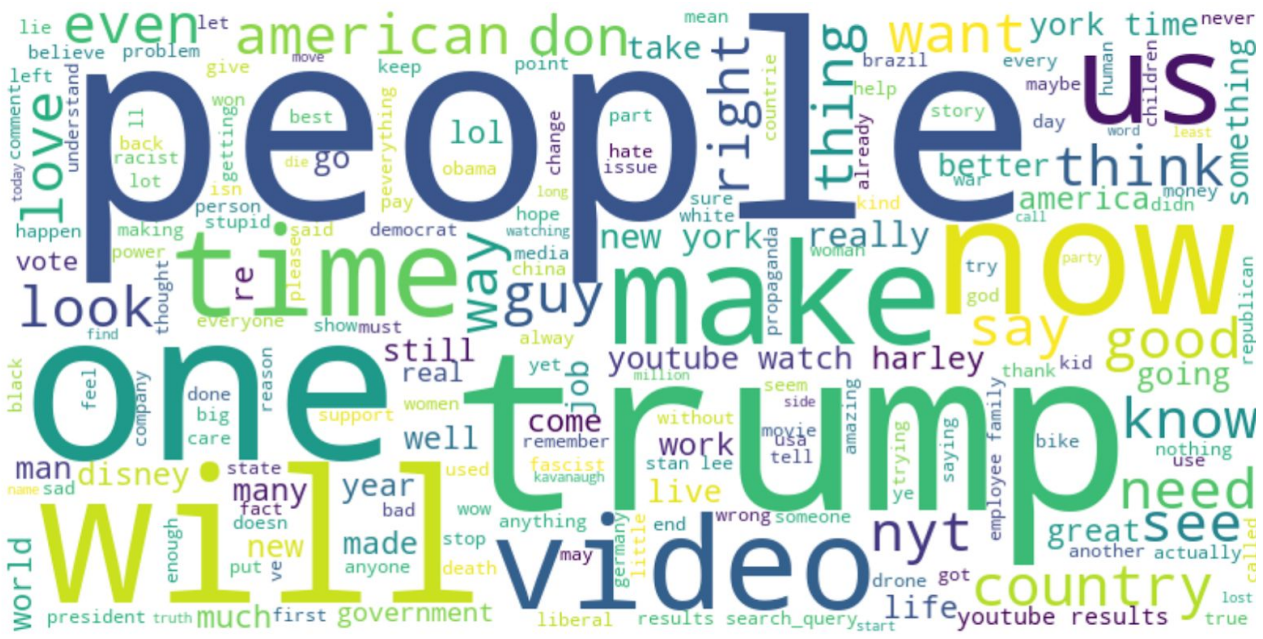

**New York Times**

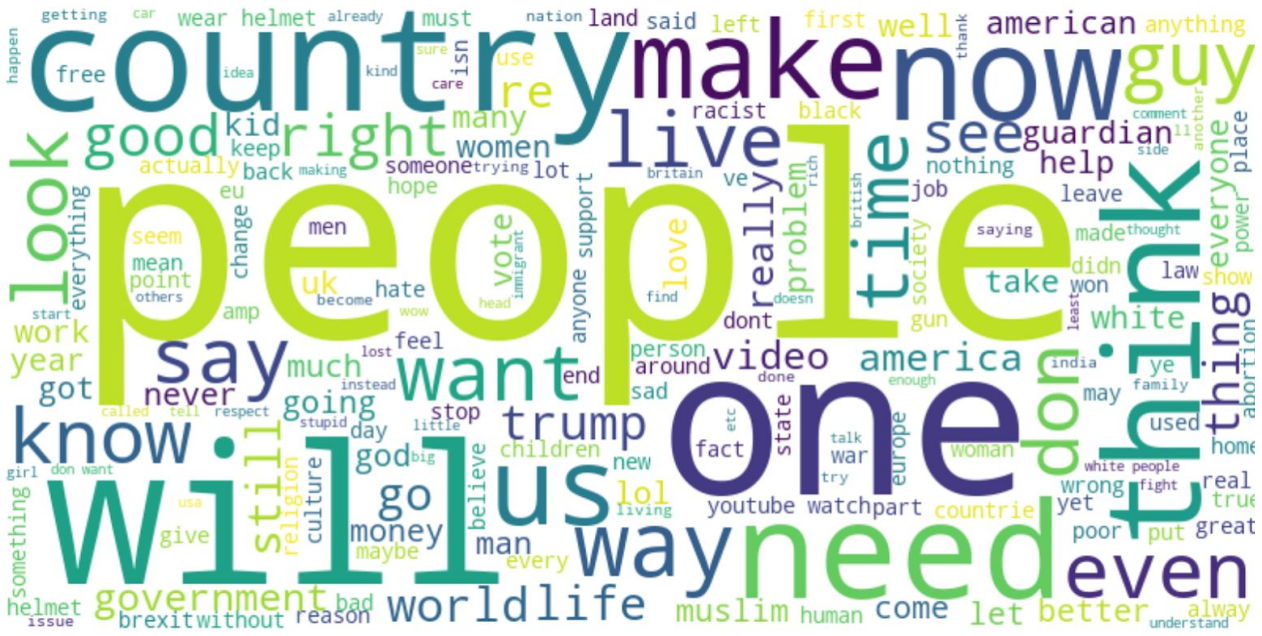

The Guardian

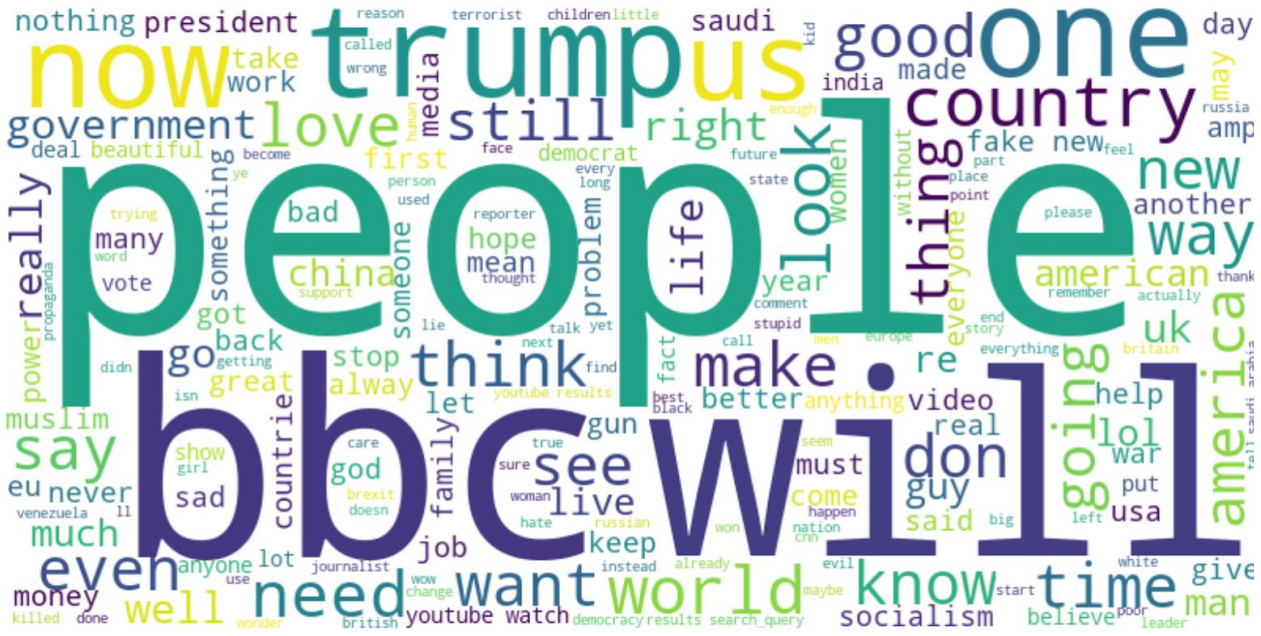

BBC

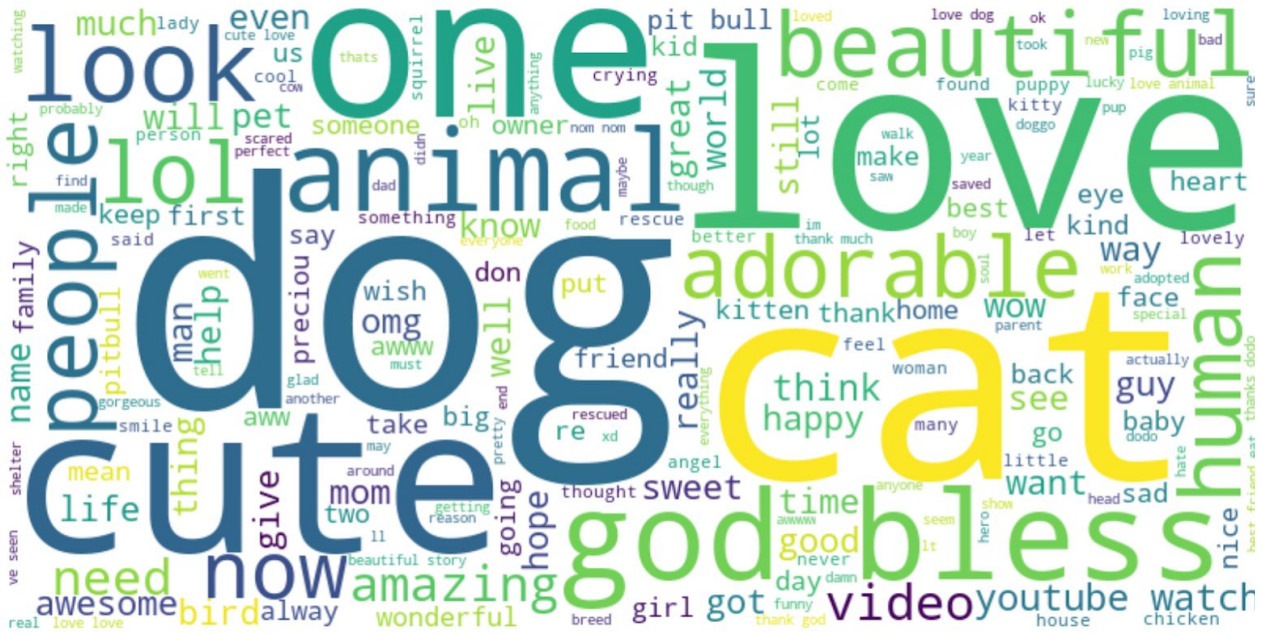

The Dodo

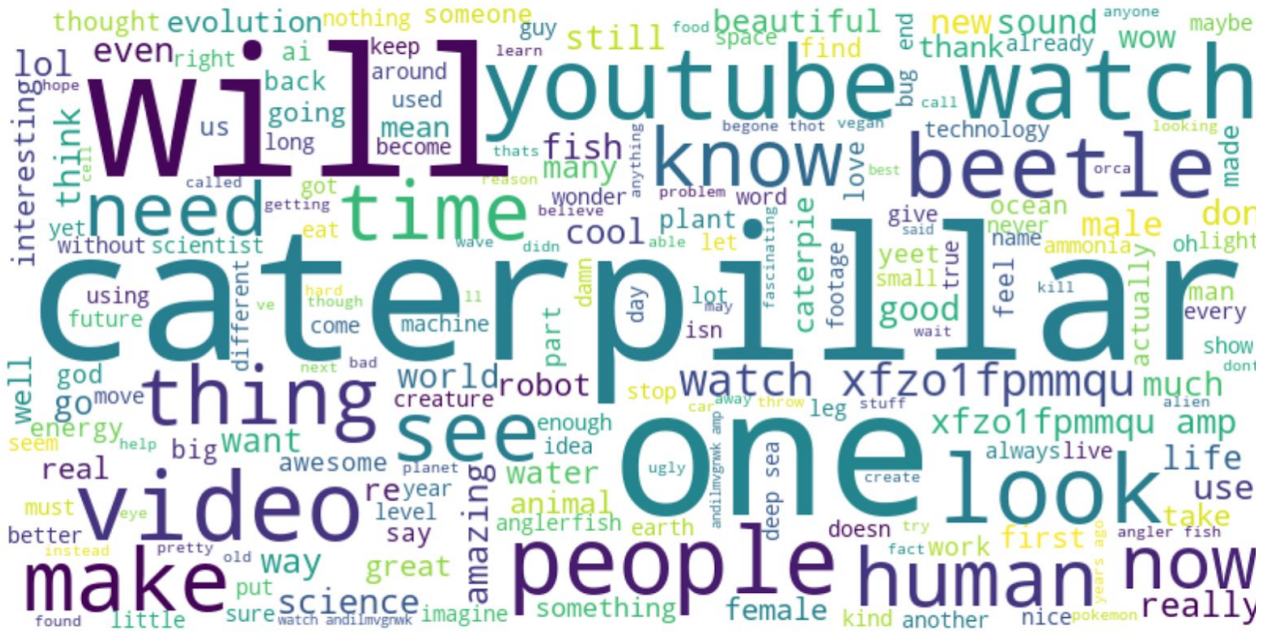

Science Magazine

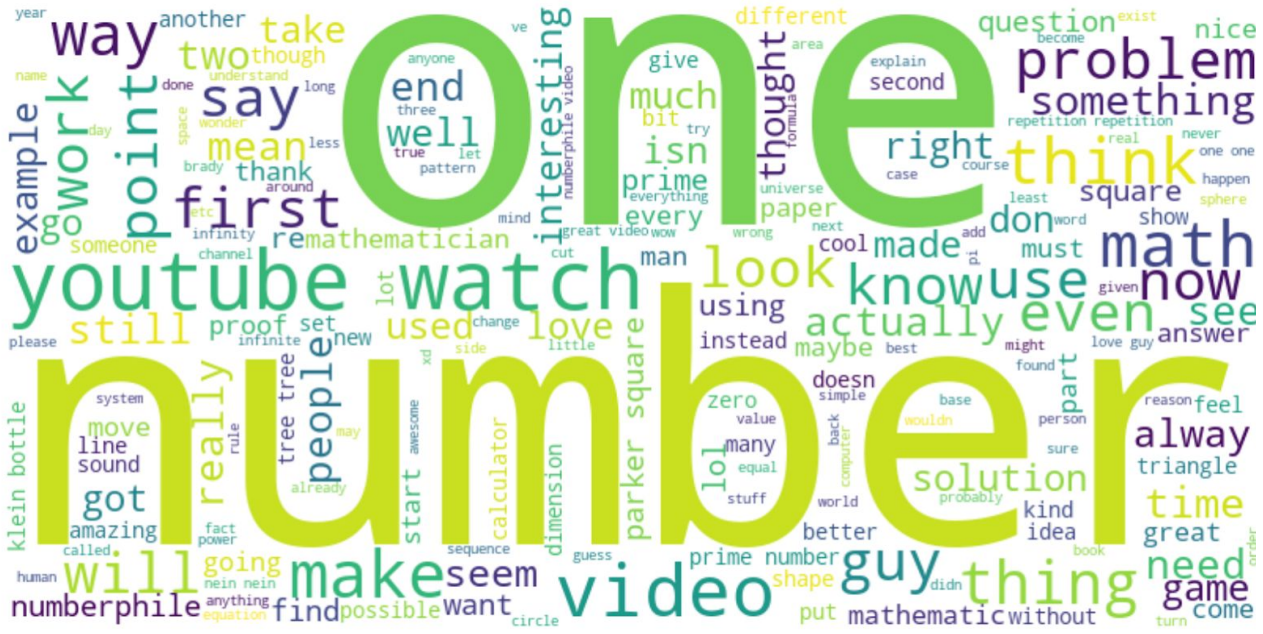

Numberphile

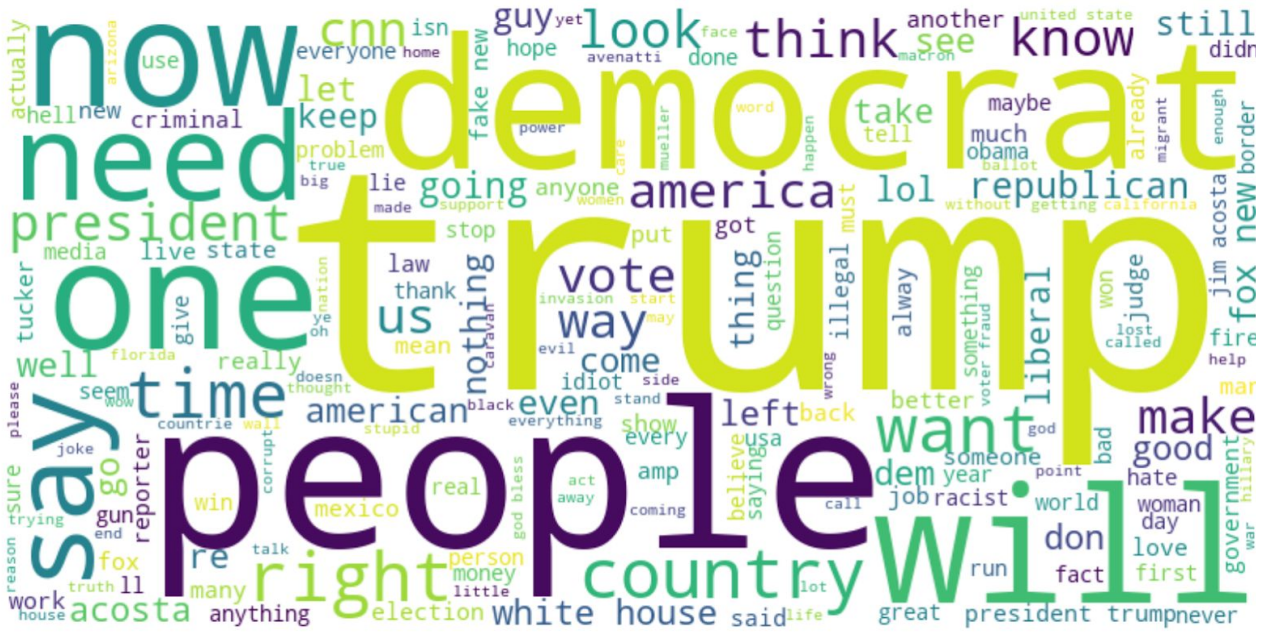

Fox News

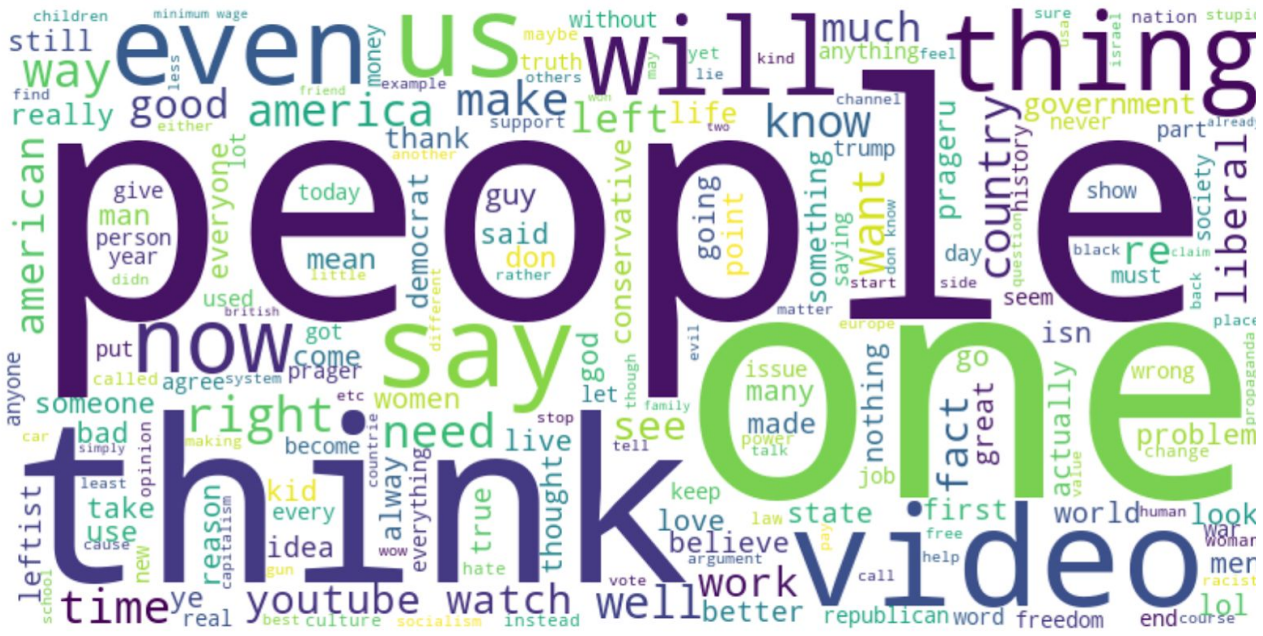

**PragerU**

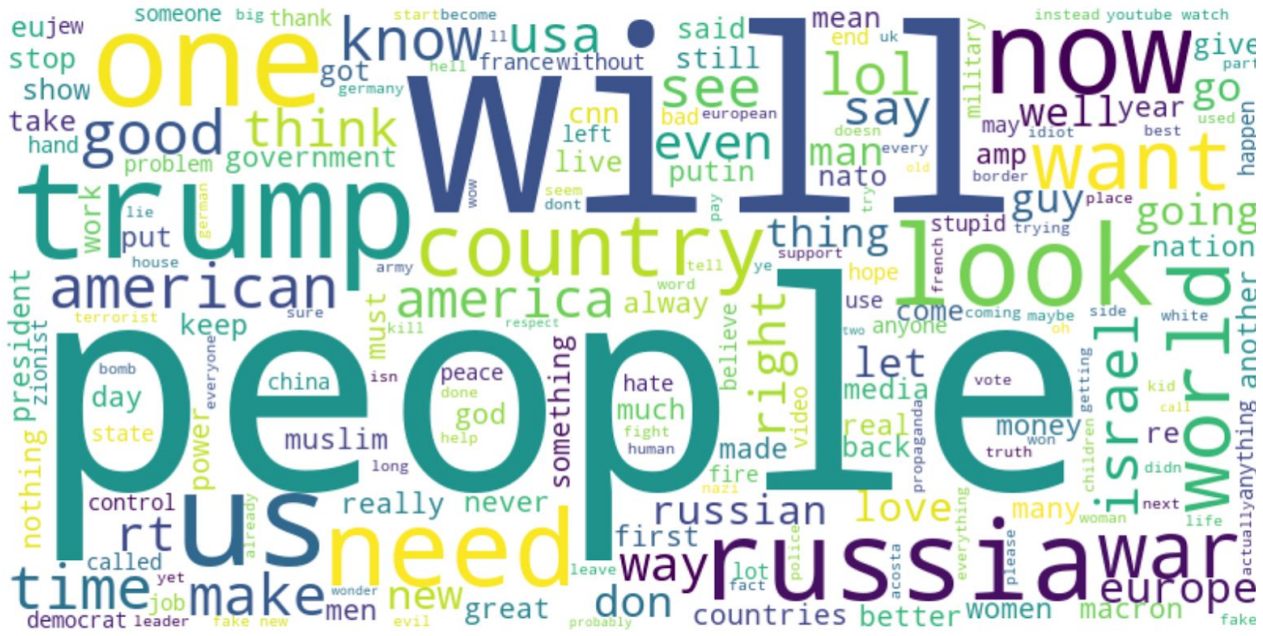

**Russia Today**

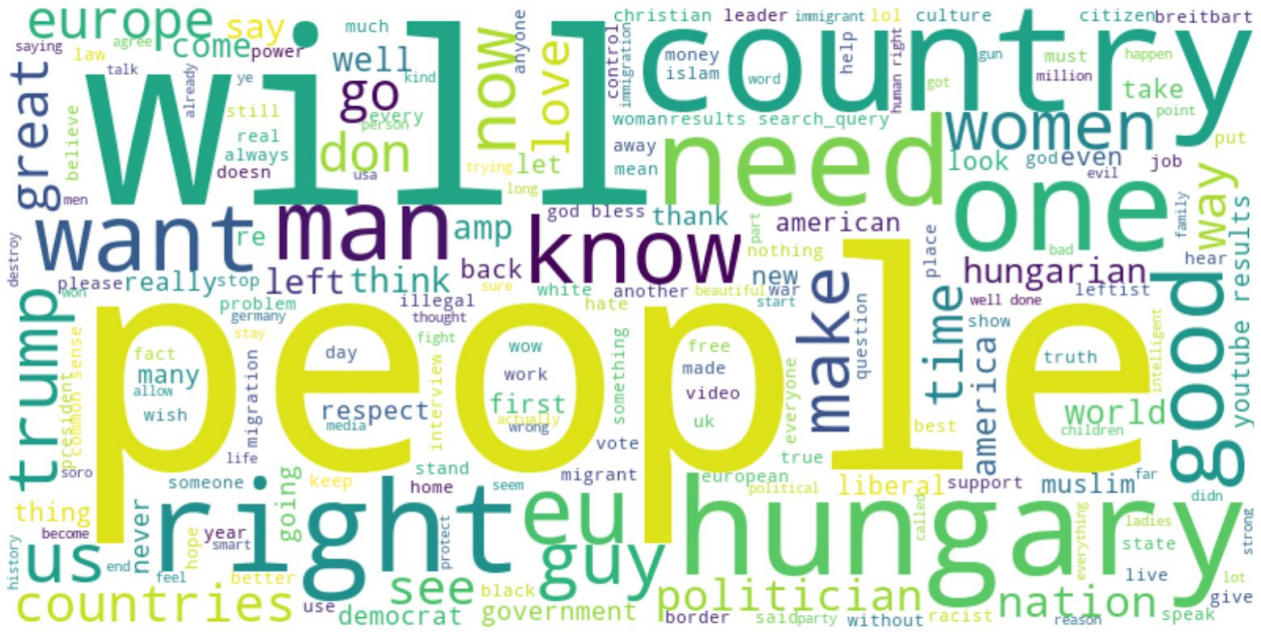

**Breitbart News**

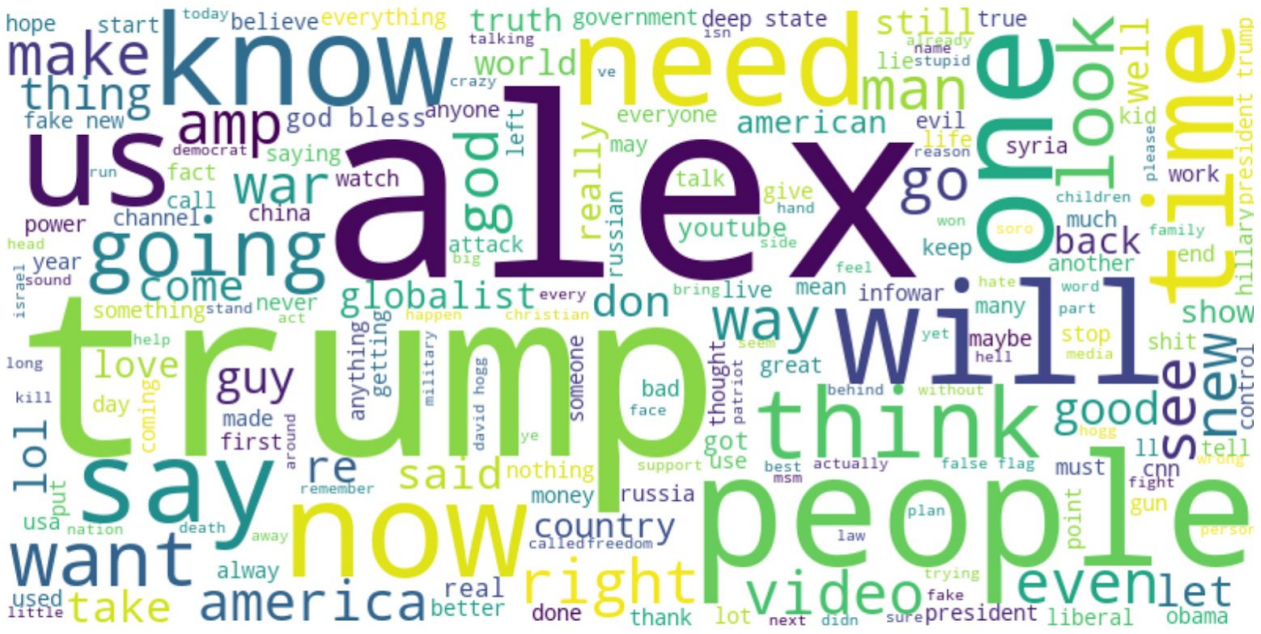

**Infowars**

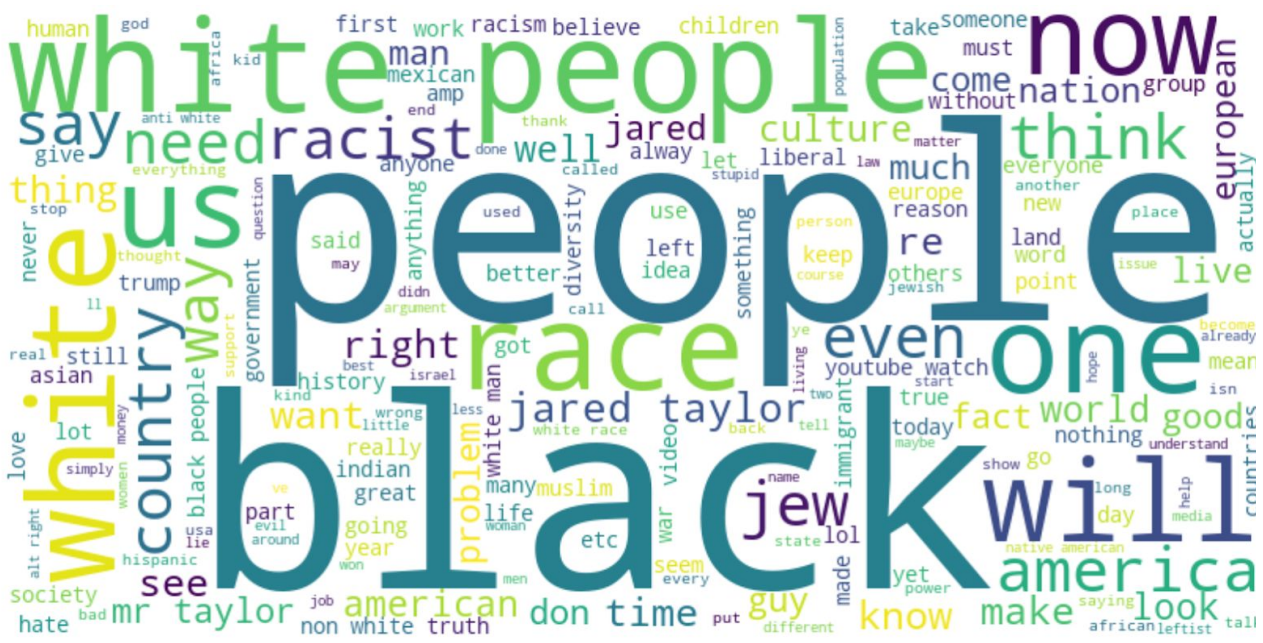

**American Renaissance**

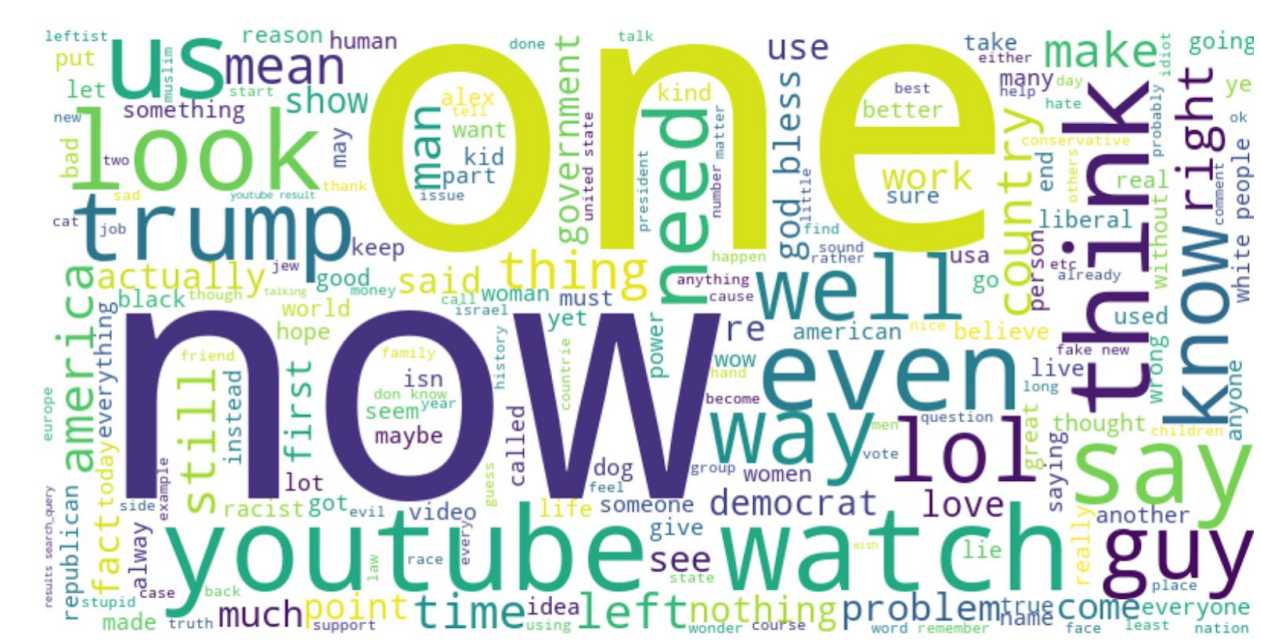

**A word cloud over the entire dataset.**

**Note**: these graphs were for an older version of our dataset, detailed in appendix item 3’s note.

### Item 5: Proportion (out of 1) of direct comments on videos on a channel with >= one reply.

While most channels have a very small % of comments with at least one reply, there are some discrepancies. Young Turks stands out with a significantly higher percentage of comments getting at least one reply, while The Dodo barely has any comments with replies (see graphic below). Young Turks videos have many short threads, whereas PragerU's threads are fewer in number but longer despite high viewership (Figure 5A). Here we see two different characterizations of comment behavior.

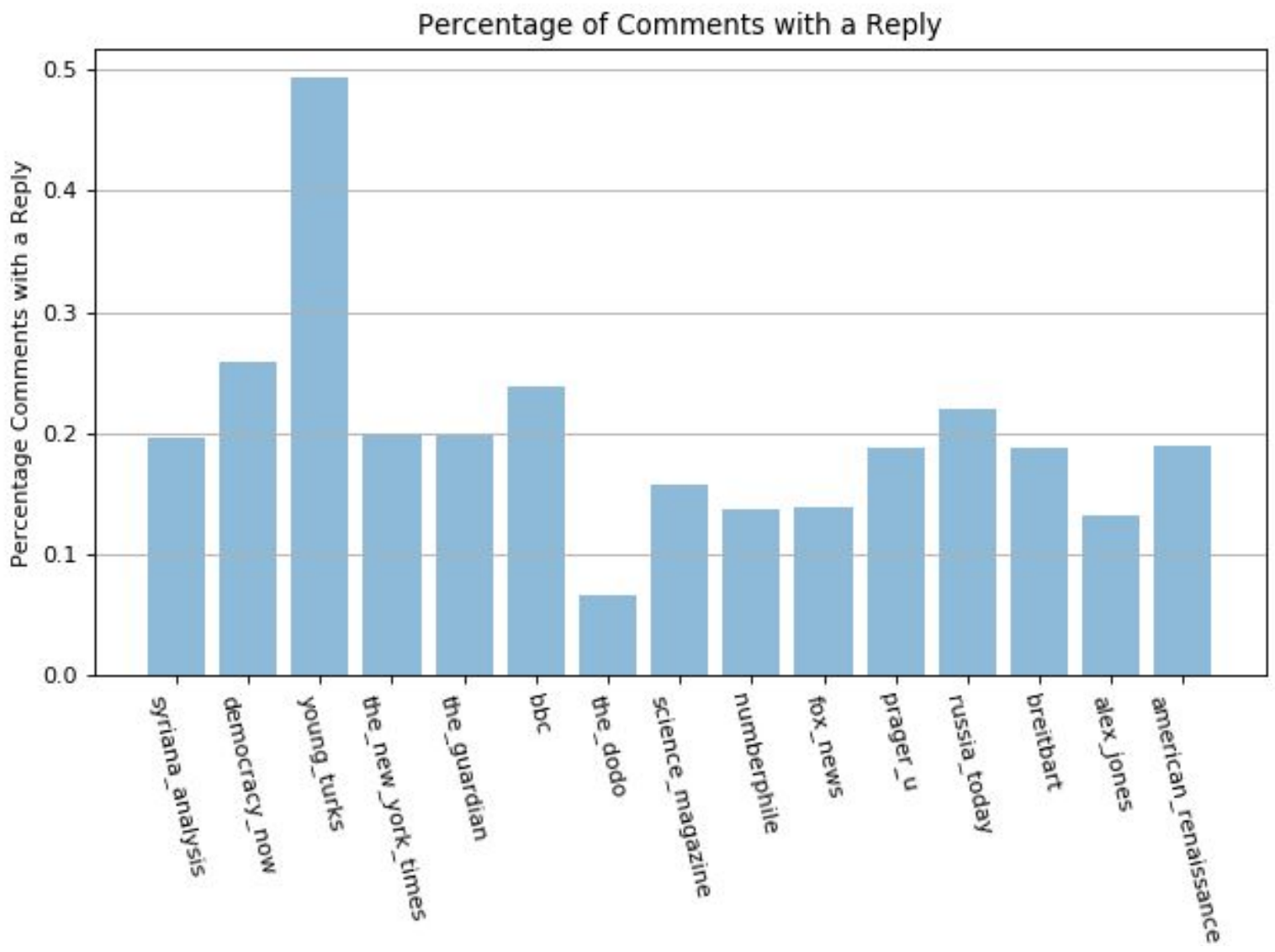

**Note**: this graph was for an older version of our dataset, detailed in appendix item 3's note.

---

### Item 6: Average thread length among all threads (comments w/ at least 1 reply) per category

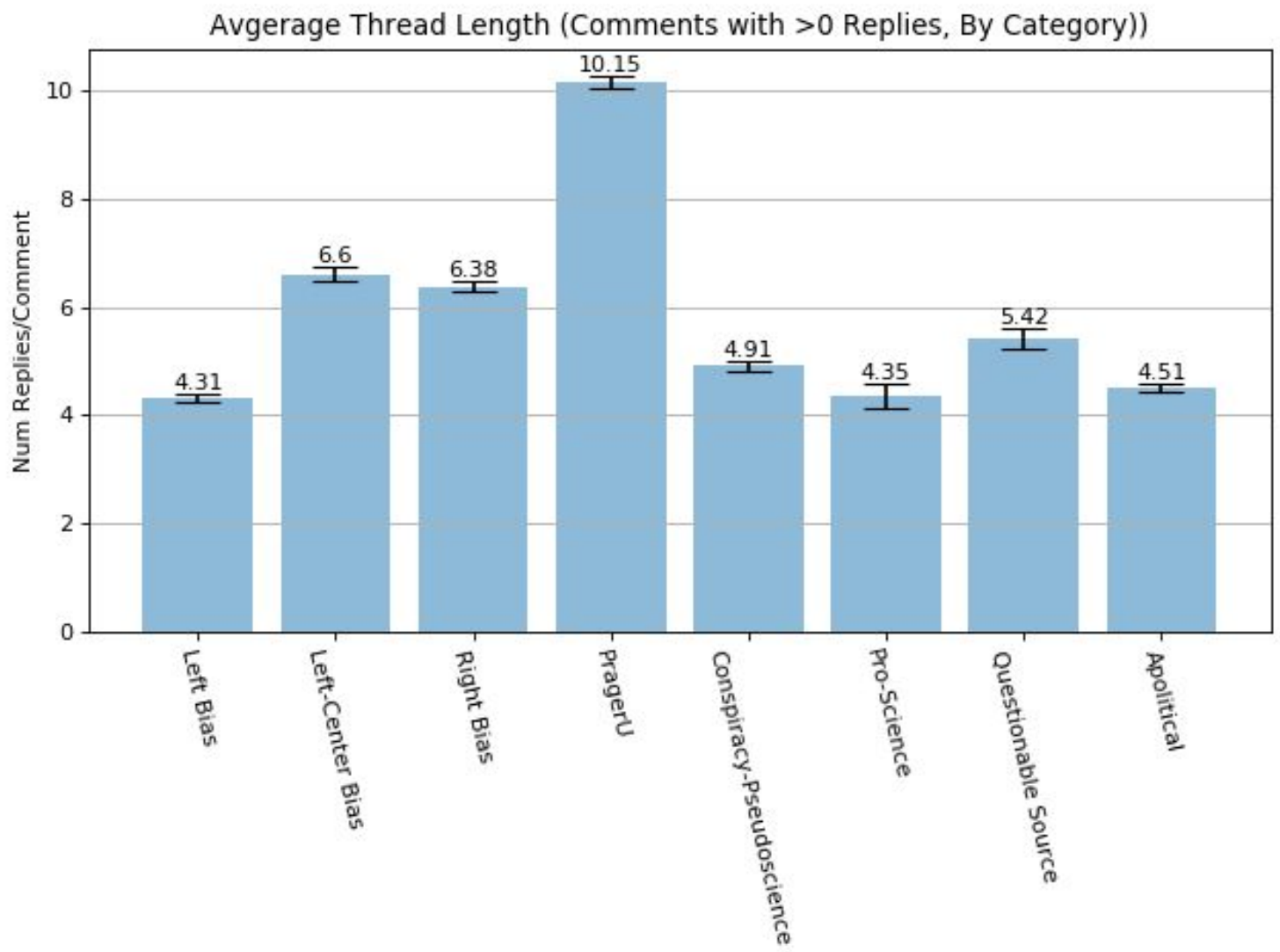

## Item 7: Breakdown of total comments and replies.

**Number of direct comments to a video, normalized by views per channel (left) and by videos per channel (right).**

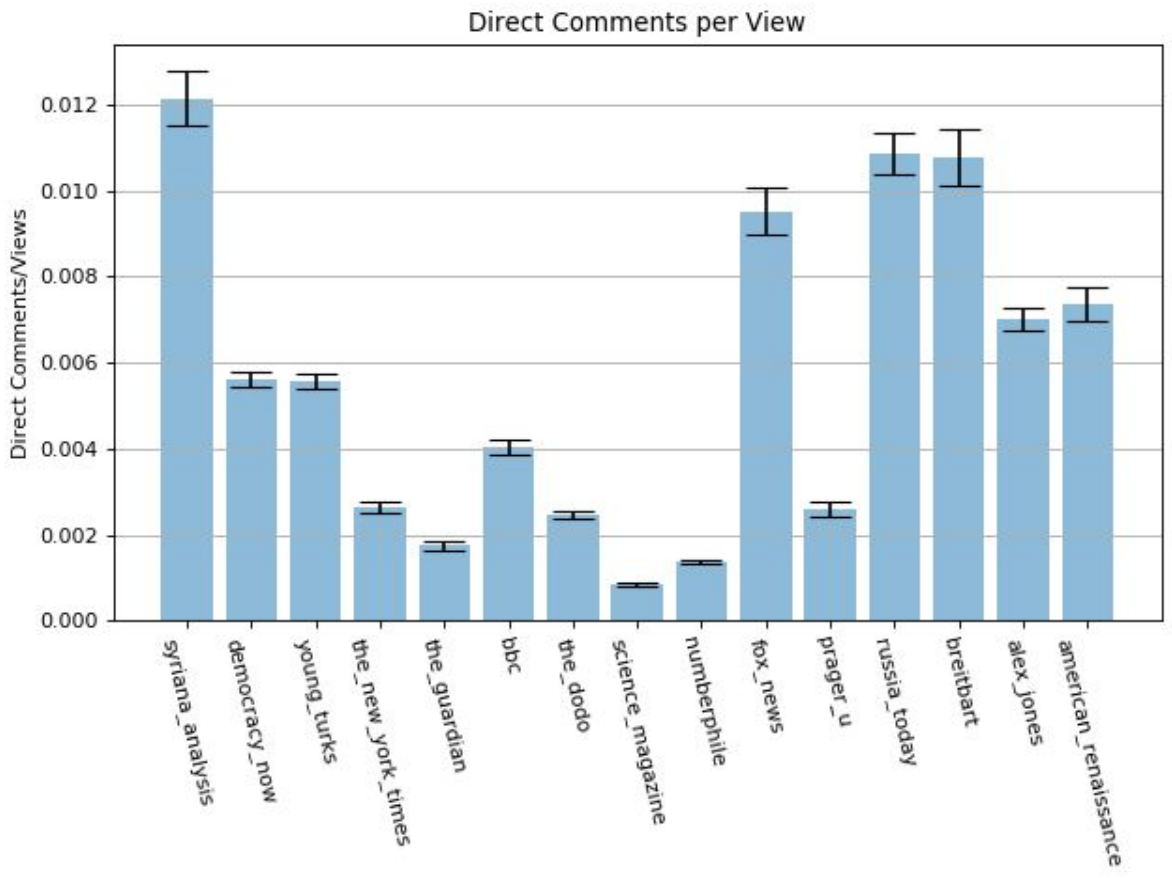

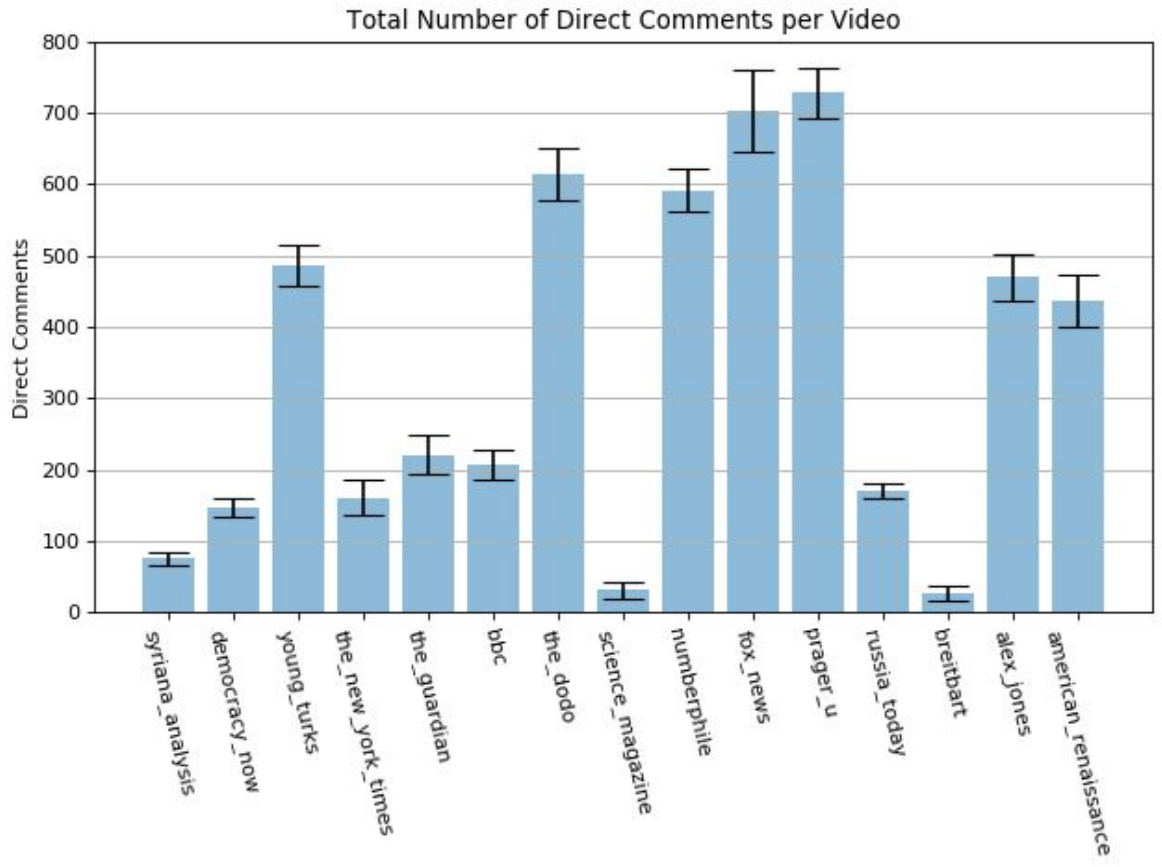

**Number of replies to a direct comment, normalized by views per channel (left) and by videos per channel (right).**

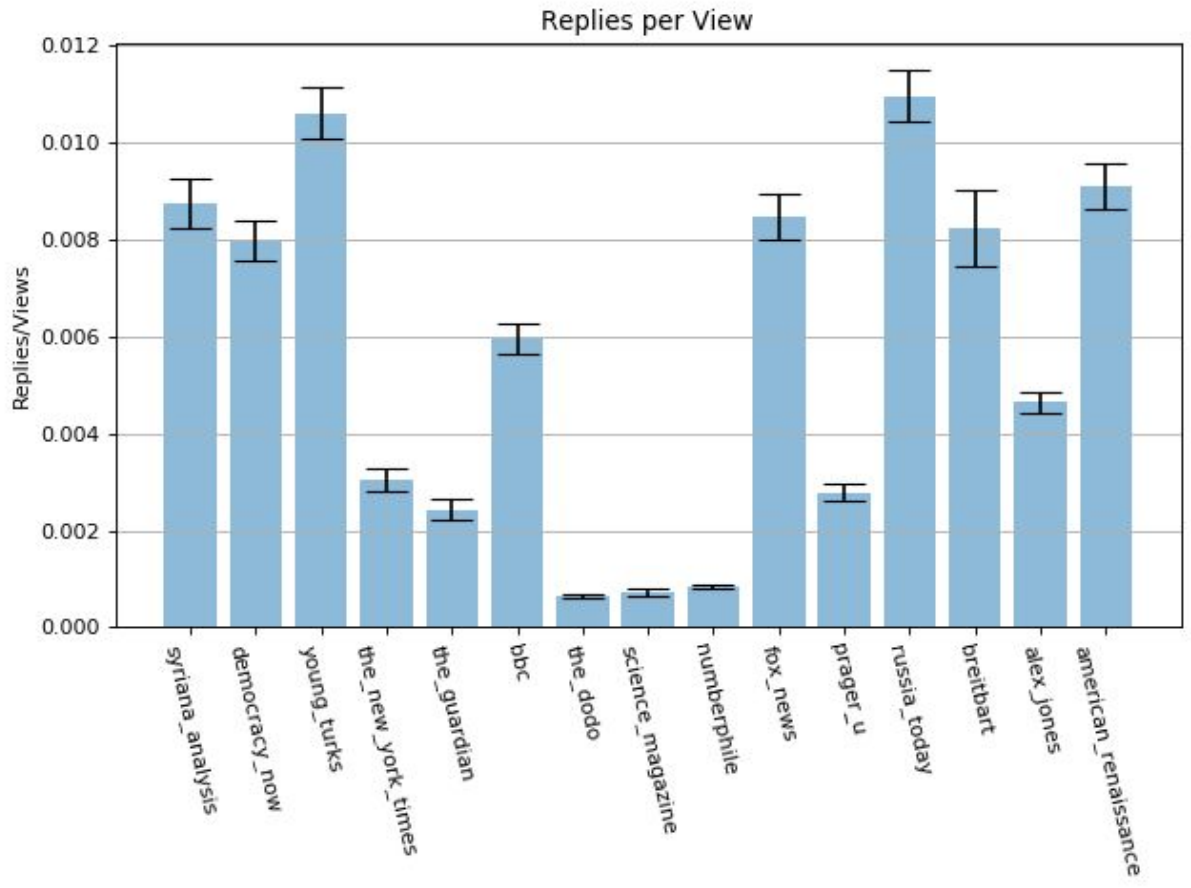

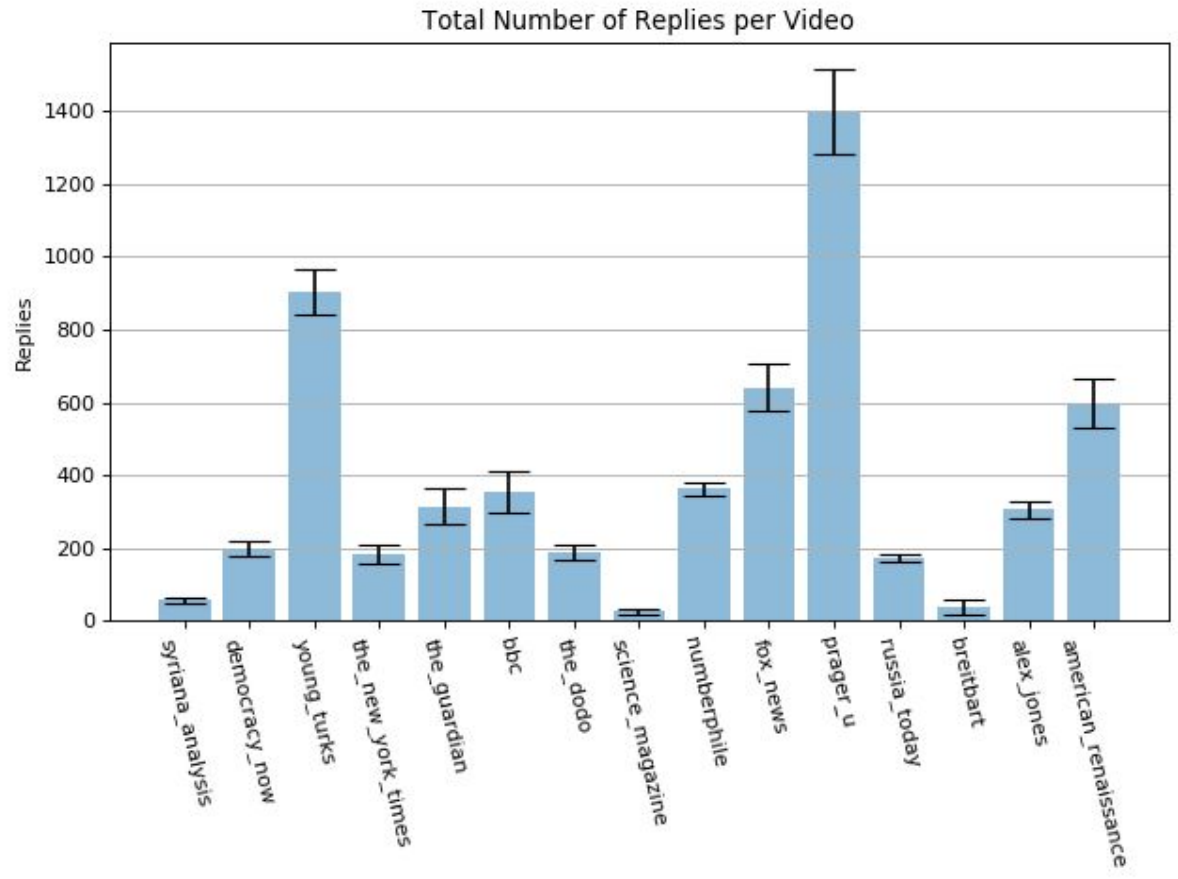

**Note**: these graphs were for an older version of our dataset, detailed in appendix item 3's note.

---

## Item 8: Overall Traction of Comments:

To establish a baseline comparison, we analyzed "like" counts and "reply" counts on comments across all channels. We find that the number of likes a comment receives is moderately correlated with the number of replies on the given comments when looking at all comments (Pearson *r* = 0.55; below graphic, titled "All Channel Comments"). This trend was observed across all YouTube channels in our dataset, but is especially present for certain right-leaning news sources such as Fox News (Pearson *r* = 0.81; below graphic, titled "fox_news_comments") and Breitbart (Pearson *r* = 0.79; below graphic, titled "breitbart_comments"). We see that the like counts and reply counts are strongly correlated for these news sources and weakly correlated across all sources. The correlation between these two measures may indicate that like count and reply count can both be used to measure the popularity of a comment as well as the engagement with a particular comment.

**Correlation between likes and number of replies on direct comments across YouTube channels. Pearson Coefficients: All $r = 0.55$, Fox News $r = 0.81$, Breitbart $r = 0.79$. All values are statistically significant, with $p \ll 0.001$. All channel correlation plots found in the appendix, item 9.**

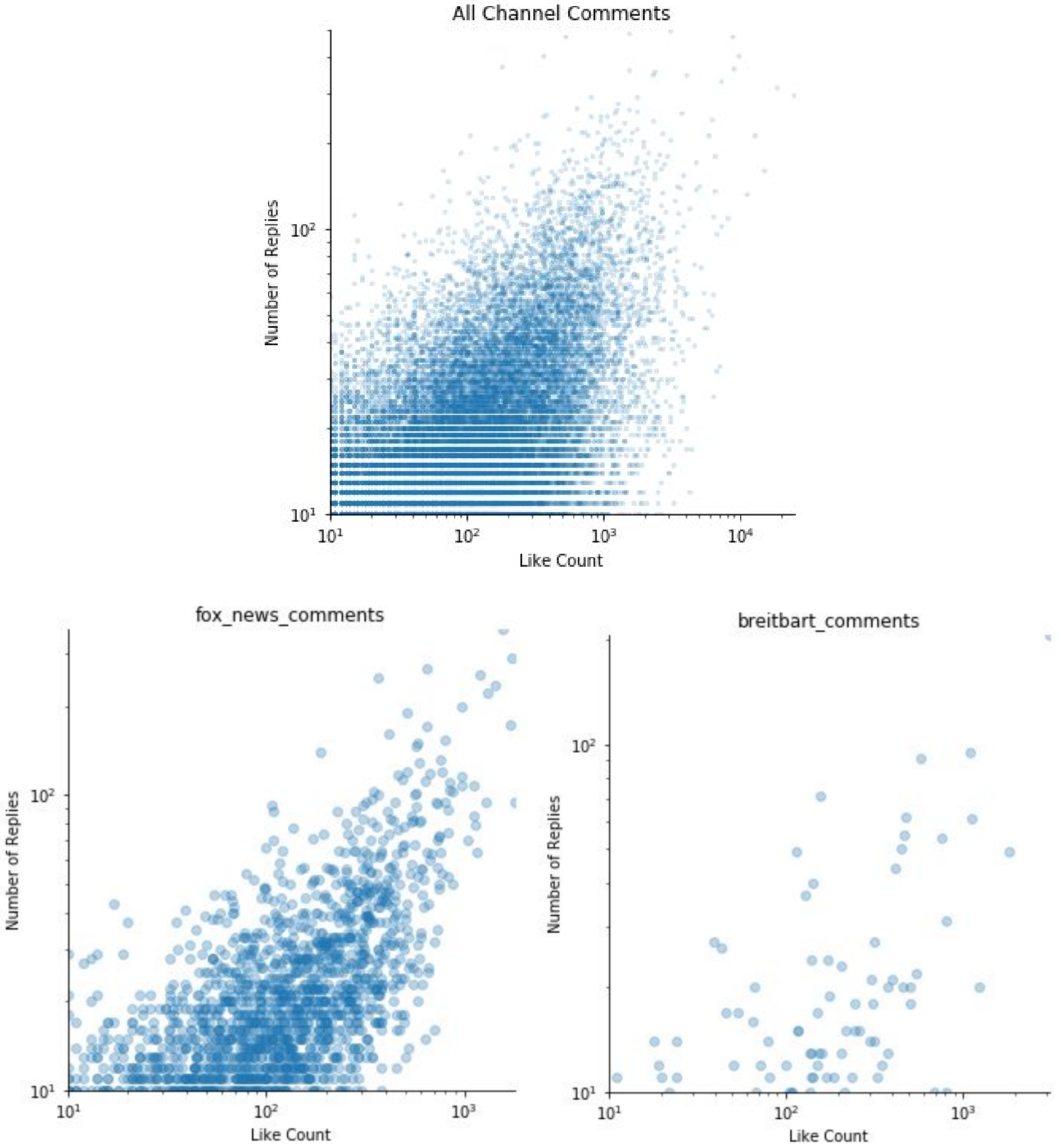

**Note**: these graphs were for an older version of our dataset, detailed in appendix item 3’s note.

## Item 9: Correlation Between Number of Replies to Comment & Number of Likes on Comment

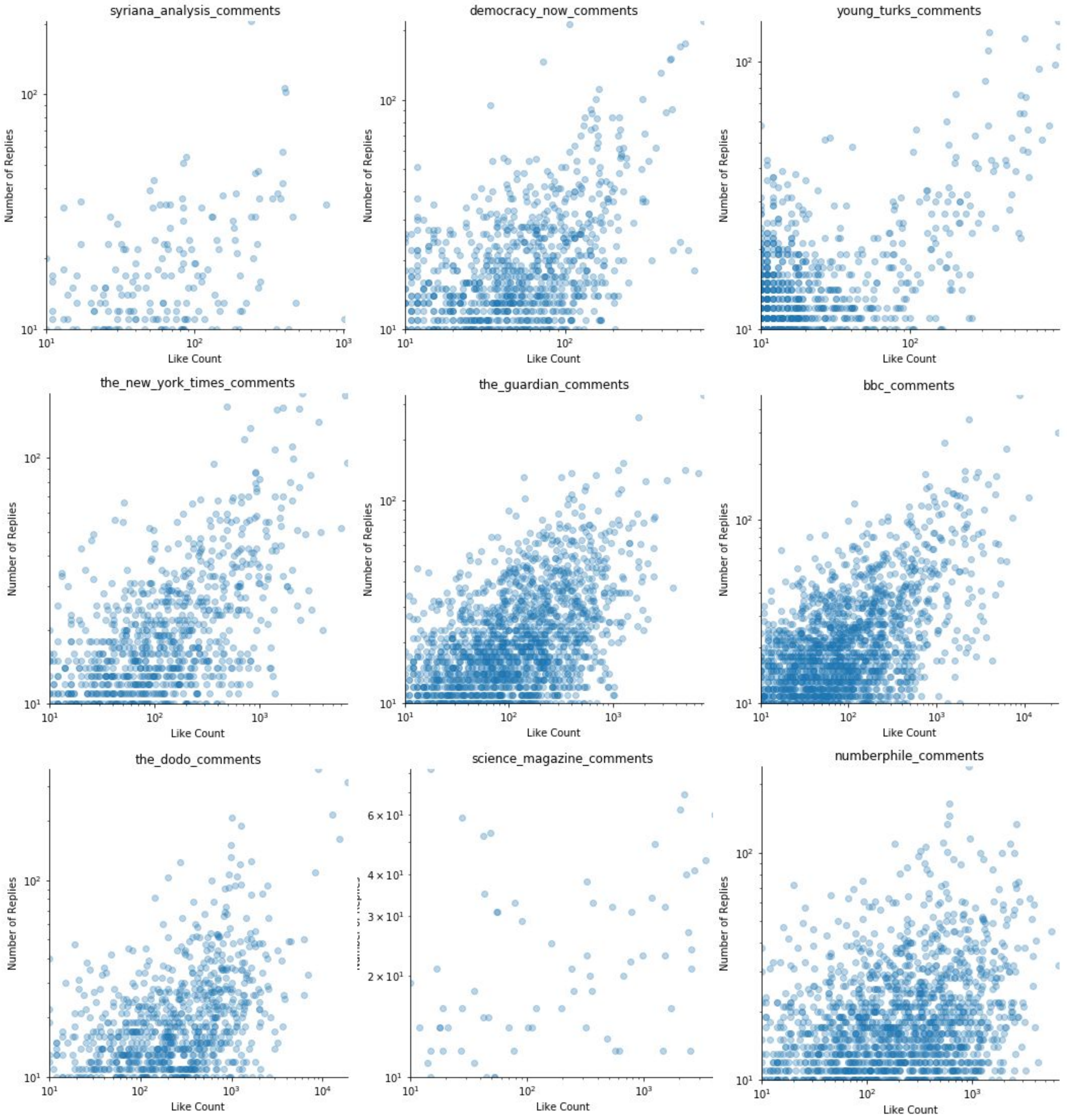

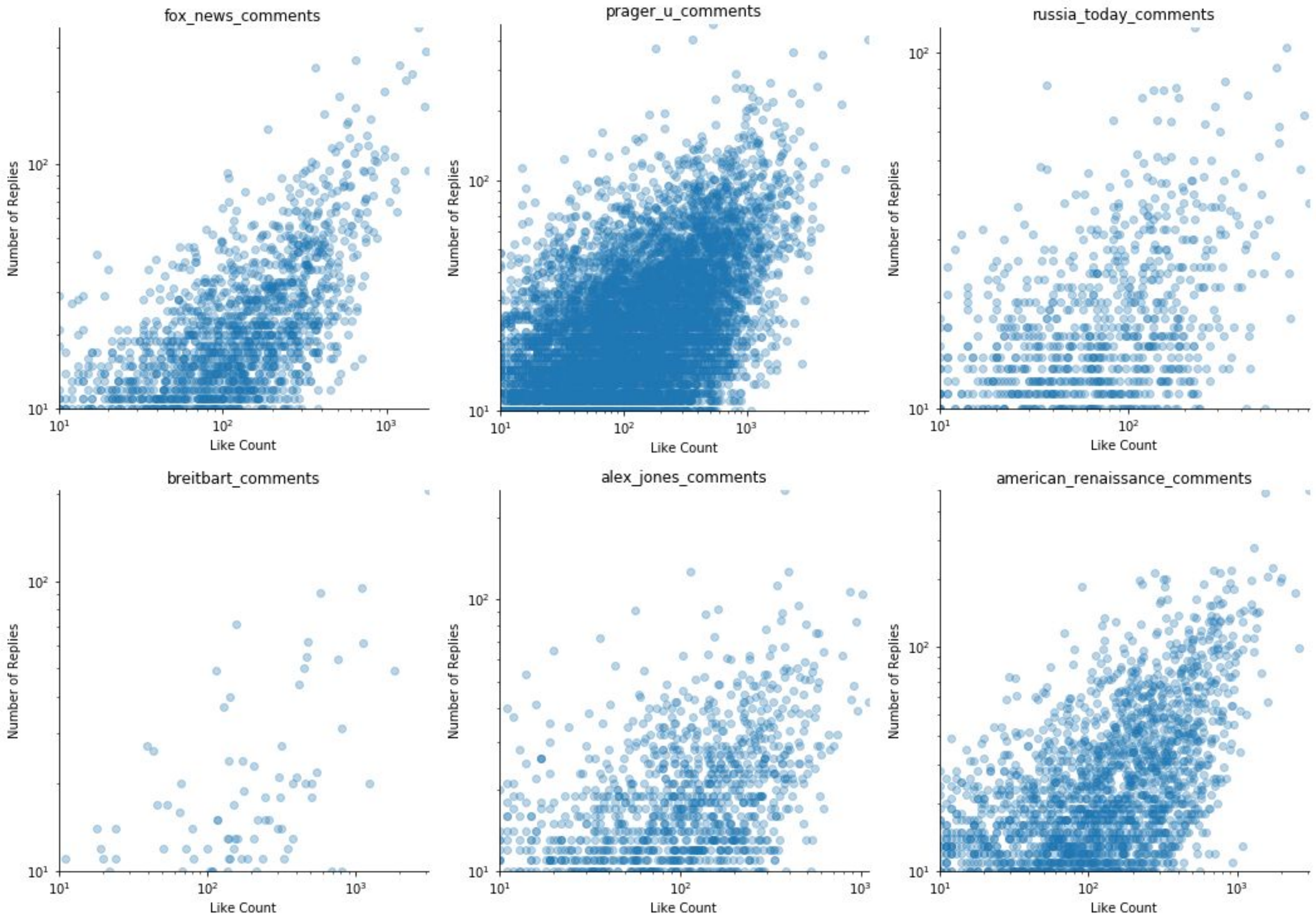

**Note**: these graphs were for an older version of our dataset, detailed in appendix item 3's note.